\documentclass[sigconf, nonacm]{acmart}

\newcommand\vldbdoi{XX.XX/XXX.XX}

\newcommand\vldbavailabilityurl{}

\usepackage{algorithm}
\usepackage[noEnd=true,indLines=true]{algpseudocodex}
\usepackage{balance}
\usepackage{cleveref}
\usepackage{subcaption}

\newcommand{\floor}[1] {\lfloor #1 \rfloor}

\newcommand{\ceil}[1] {\lceil #1 \rceil}

\newcommand{\abs}[1] {\vert#1\vert}

\begin{document}

\title{Fast and Adaptive Bulk Loading of Multidimensional Points}

\author{Moin Hussain Moti}
\orcid{0000-0002-4614-6940}
\affiliation{%
	\institution{HKUST}
	\country{Hong Kong}
}
\email{mhmoti@connect.ust.hk}

\author{Dimitris Papadias}
\orcid{}
\affiliation{%
	\institution{HKUST}
	\country{Hong Kong}
}
\email{dimitris@ust.hk}

\begin{abstract}
	Existing methods for bulk loading disk-based multidimensional points involve multiple applications of external sorting. In this paper, we propose techniques that apply linear scan, and are therefore significantly faster. The resulting FMBI Index possesses several desirable properties, including almost full and square nodes with zero overlap, and has excellent query performance. As a second contribution, we develop an adaptive version AMBI, which utilizes the query workload to build a partial index only for parts of the data space that contain query results. Finally, we extend FMBI and AMBI to parallel bulk loading and query processing in distributed systems. An extensive experimental evaluation with real datasets confirms that FMBI and AMBI clearly outperform competitors in terms of combined index construction and query processing cost, sometimes by orders of magnitude.
\end{abstract}

\maketitle

\ifdefempty{\vldbavailabilityurl}{}{
	\vspace{.3cm}
	\begingroup\small\noindent\raggedright\textbf{PVLDB Artifact Availability:}\\
	The source code, data, and/or other artifacts have been made available at \url{\vldbavailabilityurl}.
	\endgroup
}

\section{Introduction}\label{sec:intro}
In several applications (e.g., spatial/multimedia databases, recommendation systems), records can be represented as points in a multidimensional space. The coordinates of these points may correspond to locations in the Euclidean space, or the values of attributes of interest (e.g., ratings by critics/users). Queries request all records that satisfy some predicate, e.g., points within a multidimensional range, or those closest (i.e., most similar) to an input record. In the absence of an index, query processing necessitates a sequential scan of the entire data file, which is expensive due to the sheer volume of data in most applications. Multidimensional indexes enhance efficiency by pruning the parts of the data space that may not contain results. They can reside in main memory or be disk-based. Multidimensional disk-based indexes are the most common, as often the corresponding applications involve large amounts of data. They usually follow a tree structure with a single root. Internal nodes, called \emph{branches}, hold entries of their children, which can be either other branches, or \emph{leaves}, at the bottom tree level. Each node has a spatial extent, often shaped as a $d$-dimensional hyper-rectangle, where $d$ is the dimensionality, that covers all points within its sub-tree.

Ideally, a multidimensional index should exhibit several, sometimes conflicting, characteristics.
(1) It should be fast to build and update.
(2) It should be space efficient, packing nodes close to their maximum capacity because half empty nodes have a negative effect on queries with large output.
(3) It should minimize the total node area per level.
(4) It should have shapely (i.e., square-like) node extents, avoiding nodes that are elongated on some dimension. Such nodes are likely to be accessed by many queries, even though they may not contain results.
(5) It should minimize overlapping nodes at the same level because all such nodes are visited by queries that intersect the overlapping area.

When the data are given in advance, various bulk loading methods generate disk-based indexes using \emph{external sorting}, as opposed to a distinct insertion per record. These methods differ on the order of level creation (\emph{top-down} methods generate the root node first, while \emph{bottom-up} create the leaf level), and the resulting index type. Conventional bulk loading creates the entire index in a single step. On the other hand, \emph{adaptive indexes} are built progressively as a response to query processing. Consequently, parts of the index that contribute query results, are more refined than the rest. Although there has been some recent work on adaptive bulk loading for multidimensional points, it is focused on main memory.

In the following, we propose a novel bulk loading method for disk-based multidimensional points, which relies on scanning, as opposed to external sorting, and is much faster than existing techniques. The resulting index, called FMBI (Fast Multidimensional Bulkloaded Index) is very efficient for query processing as it exhibits the above desirable characteristics, including almost full and square nodes with zero overlap. In addition, we extend the proposed techniques to derive an adaptive version called AMBI (Adaptive Multidimensional Bulkloaded Index) that is refined using the query workload, gradually transforming to the final index. AMBI has a huge advantage compared to non-adaptive competitors, when the queries cover a small part of the data space, in which case only a partial index is generated. Our contributions are:

\begin{enumerate}
	\item Novel scan-based techniques for bulk loading disk resident multidimensional points.
	\item FMBI, a multidimensional index that exhibits excellent query performance, while it is several times faster to build than its competitors.
	\item An adaptive version AMBI that builds the index on-demand according to the query workload, and avoids unnecessary work for parts of the data space that do not contribute query results.
	\item Parallel versions of FMBI and AMBI, suitable for distributed systems and spatial partitioning.
	\item A comprehensive experimental evaluation with real datasets that compares FMBI and AMBI with a multitude of existing indexes under a unified framework.
\end{enumerate}

The rest of the paper is organized as follows. \Cref{sec:related-work} presents related work. \Cref{sec:fmbi} describes the proposed bulk loading algorithms and FMBI. \Cref{sec:ambi} extends our work to adaptive indexing and AMBI. \Cref{sec:parallel} discusses parallel bulk loading in distributed systems. \Cref{sec:exp} contains the experimental evaluation, and \Cref{sec:conclusion} concludes the paper.

\section{Related Work}\label{sec:related-work}
This section contains background material on multidimensional bulk loading and related areas.

\begin{figure*}[t]
	\centering
	\begin{subfigure}{0.49\textwidth}
		\includegraphics[width=0.98\textwidth]{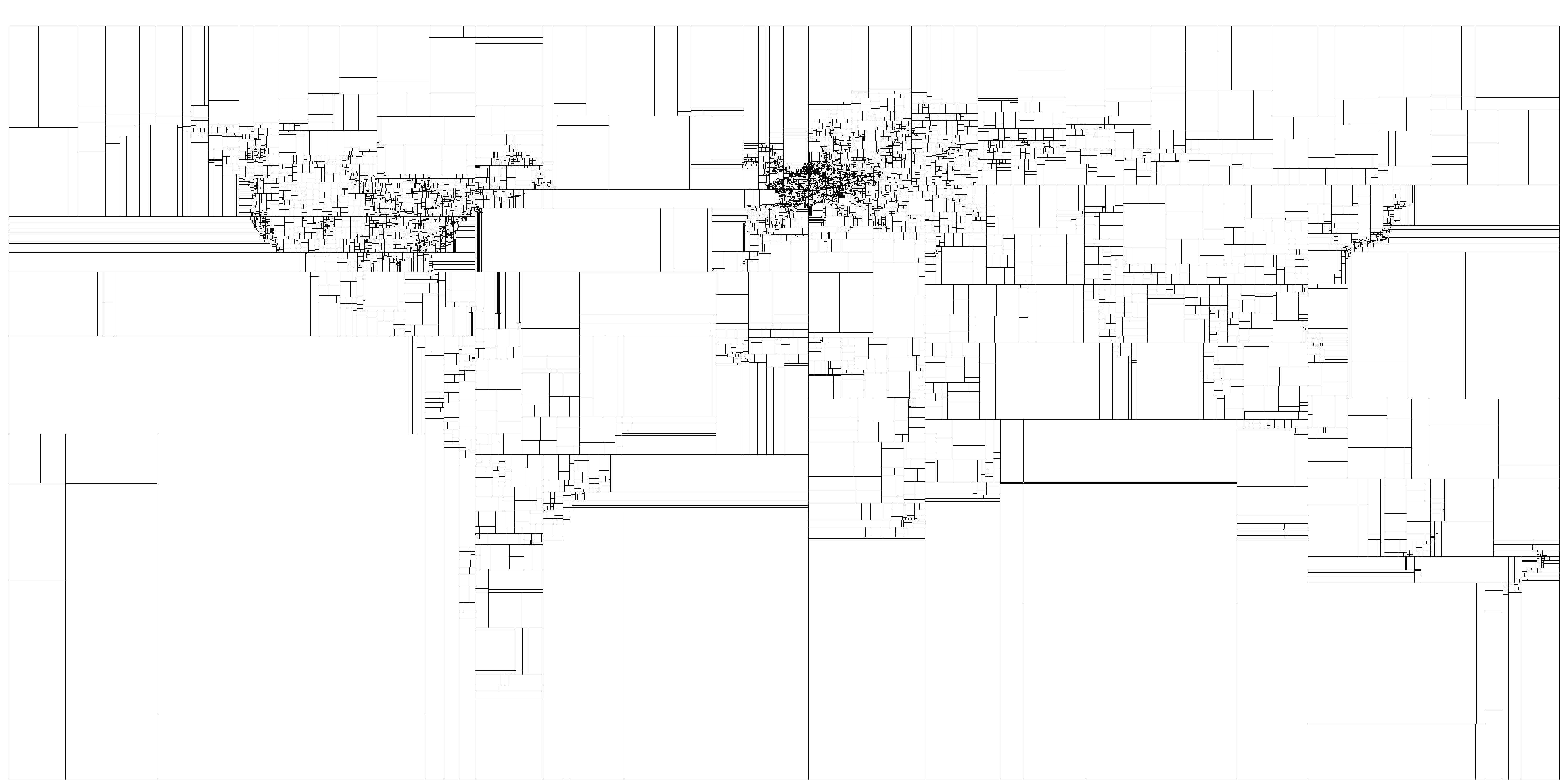}
		\caption{Spread KDB-Tree}
		\label{fig:kdb_spread}
	\end{subfigure}
	\begin{subfigure}{0.49\textwidth}
		\includegraphics[width=0.98\textwidth]{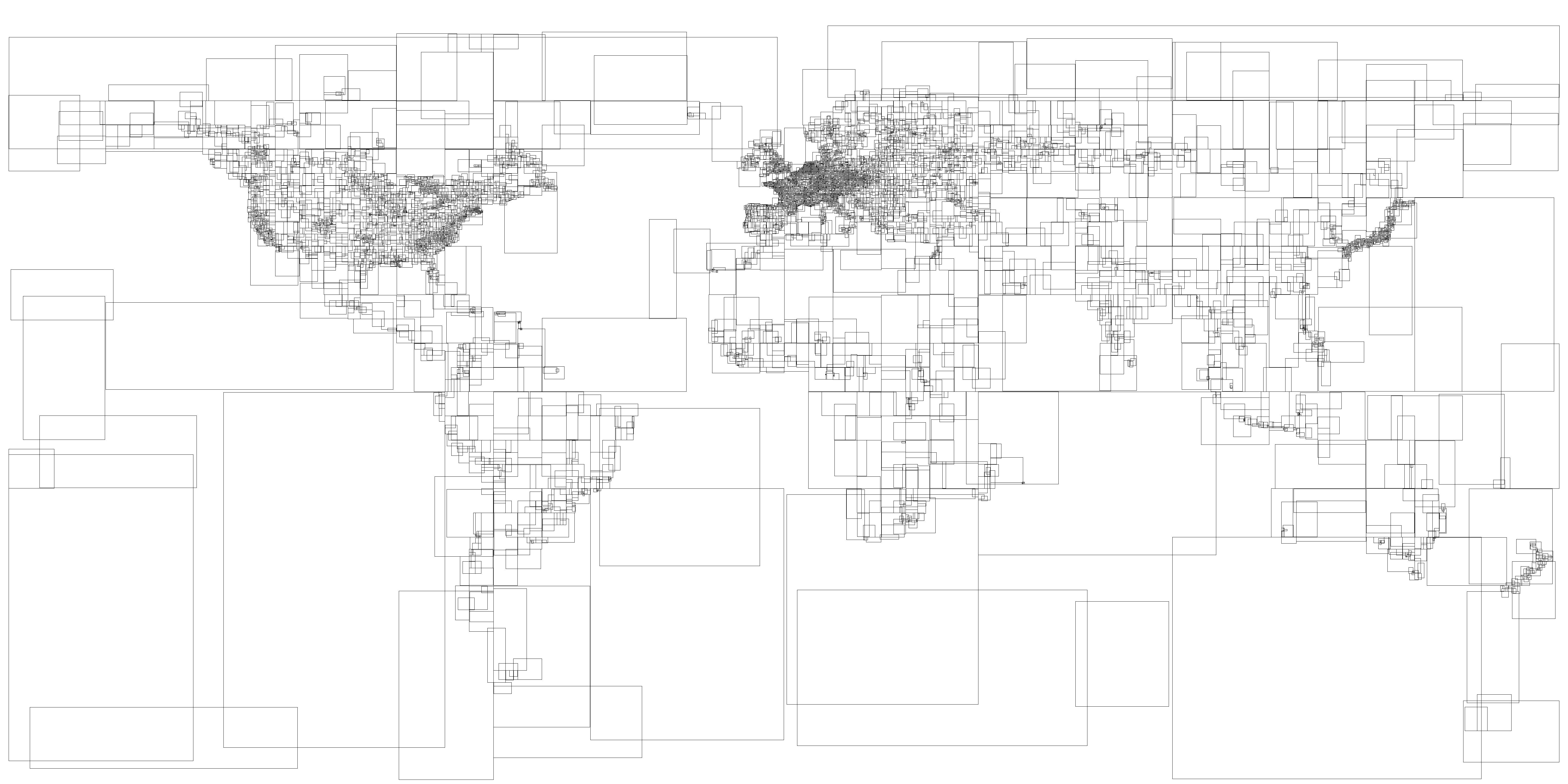}
		\caption{Hilbert}
		\label{fig:hilbert}
	\end{subfigure}
	\begin{subfigure}{0.49\textwidth}
		\includegraphics[width=0.98\textwidth]{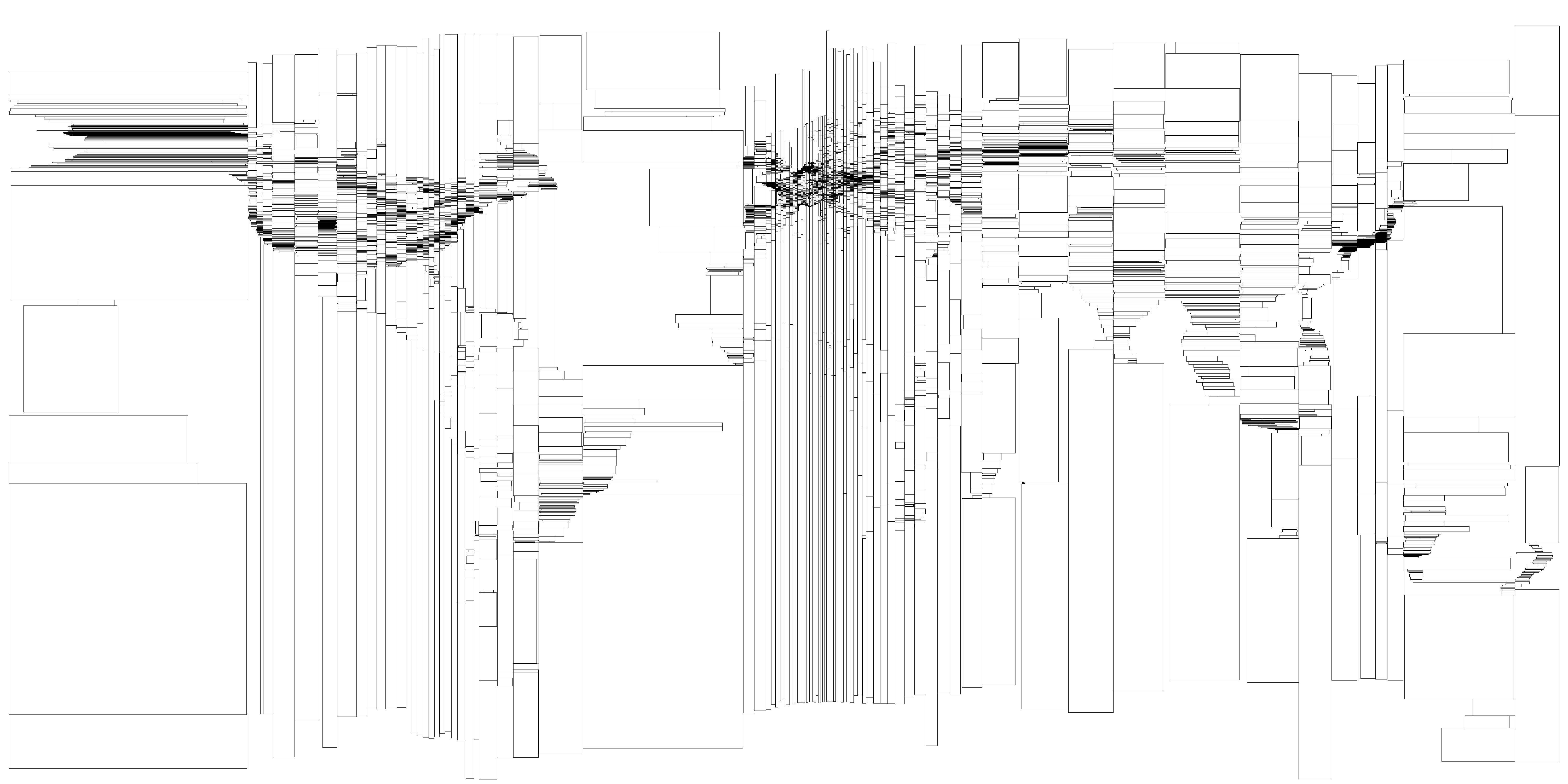}
		\caption{STR}
		\label{fig:str}
	\end{subfigure}
	\begin{subfigure}{0.49\textwidth}
		\includegraphics[width=0.98\textwidth]{figures/snapshots/TGS.pdf}
		\caption{TGS}
		\label{fig:tgs}
	\end{subfigure}
	\begin{subfigure}{0.49\textwidth}
		\includegraphics[width=0.98\textwidth]{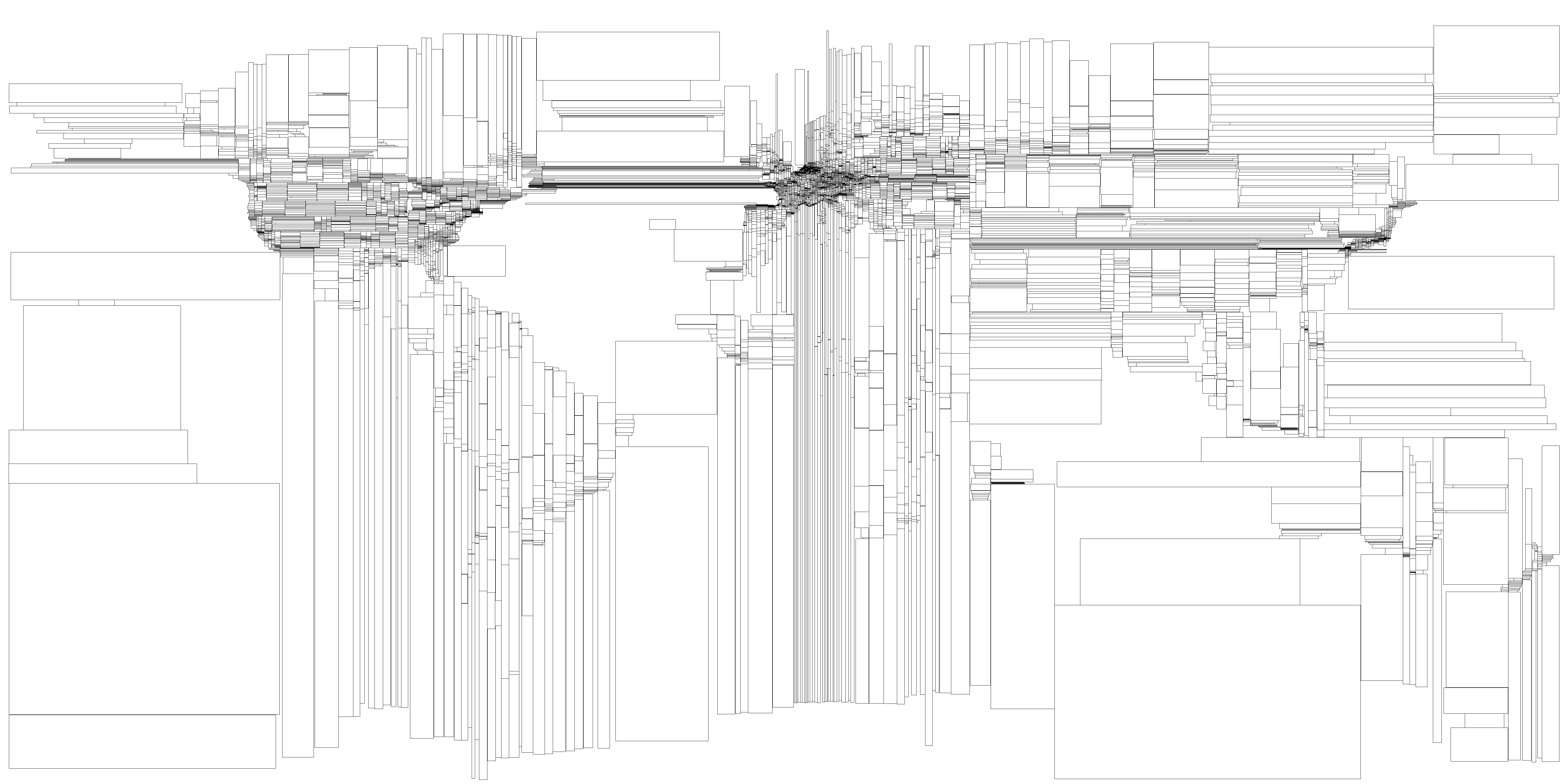}
		\caption{OMT}
		\label{fig:omt}
	\end{subfigure}
	\begin{subfigure}{0.49\textwidth}
		\includegraphics[width=0.98\textwidth]{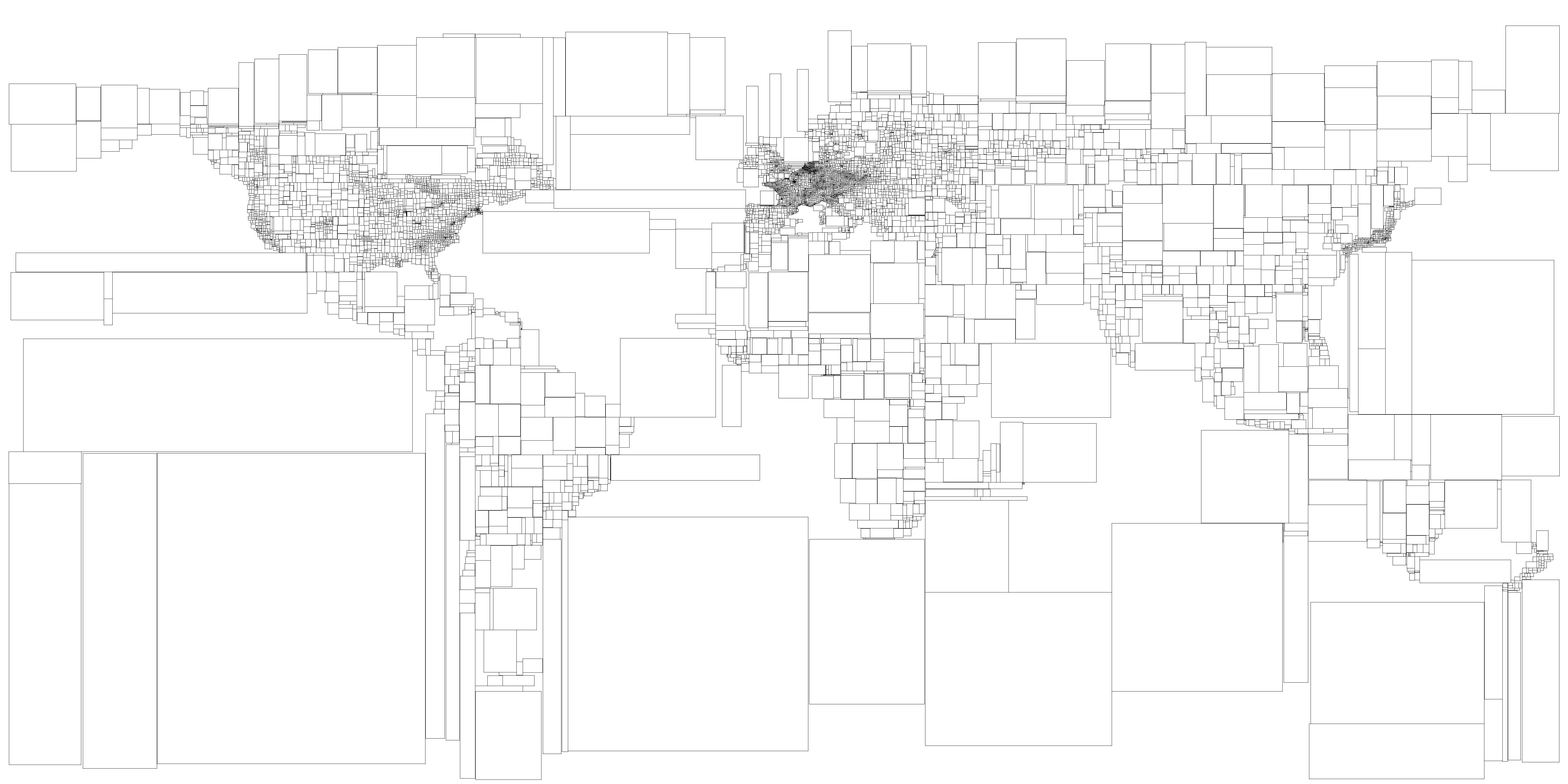}
		\caption{Waffle}
		\label{fig:waffle}
	\end{subfigure}
	\caption{OSM leaf nodes created by bulk loading methods}
	\label{fig:snapshots}
\end{figure*}

\subsection{Multidimensional Bulk Loading Methods}\label{sec:multidimensional-bulk-loading-methods}
Multidimensional indexes are based on \emph{space}, or \emph{data} partitioning. In space partitioning schemes (e.g., KDB-Trees), the set of nodes at every level covers the entire data space. In data partitioning (e.g., R-trees), a node's spatial extent is the minimum bounding box (MBBs) covering its entries, so that there may exist empty spaces not covered by any node. Various methods bulk load both index types starting from the root (top-down) or the leaf level (bottom-up).


KDB-Trees ~\cite{robinson_kdbcyclic_1981} are considered the most effective space partitioning index. They can be bulk loaded using a top-down process ~\cite{moti_waffle_2022}, which begins with a single node holding all the pages and covering the entire space. Partitioning is performed by sorting the pages on a selected dimension, and splitting on the median entry. If the resulting subspaces overflow, they are recursively split. KDB-Tree variants are differentiated on the choice of split dimension.
\emph{Cyclic} KDB-Trees~\cite{robinson_kdbcyclic_1981} alternate the split dimension, while \emph{Spread} KDB-Trees~\cite{friedman_kdbspread_1977} select the longest dimension, i.e., the one with the highest data-spread.

A number of methods bulk load R-trees bottom-up. In Hilbert Packing~\cite{kamel_hilbert_1993} the set of data points is first sorted on their Hilbert rank~\cite{fisher_hilbert_1986}. Then, groups of $C_L$ consecutive sorted points, where $C_L$ is the leaf node capacity, are packed into disk pages, which form the bottom level (i.e., leaf nodes). The same procedure is applied to the nodes of the next level, all the way to the root, using a corner or center of a node's spatial extent to compute its Hilbert rank. Essentially this method generates an 1D order (based on the Hilbert rank) of multidimensional objects. Alternative techniques can be based on different space filling curves \cite{achakeev_query_adaptive_2012}~\cite{qi_packing_2020}, or simply on sorting on a single dimension \cite{roussopoulos_pictorial_1985}. Such methods achieve full disk pages, but may lead to overlapping nodes.

STR~\cite{leutenegger_str_1997} is a bottom-up procedure based on sorting and tiling. Given a 2-dimensional dataset of $N$ points, the number of pages required for the leaf level is $P = \ceil{N/C_L}$. The dataset is sorted on the $x$-axis, and the points are \emph{tiled} to $\ceil{\sqrt{P}}$ vertical slices. Slices with more than $C_L$ points are sorted and tiled on the $y$-axis. The slices that remain after all overflows have been resolved form the leaf level of the R-tree. The process is repeated recursively for the resulting nodes to create subsequent levels, until there is a single root. STR can be adapted to $d$-dimensional ($d>2$) datasets by tiling $\ceil{P^{\frac{1}{d}}}$ slices for each dimension.

TGS~\cite{garcia_tgs_1998} is a top-down bulk loading procedure for the R-Trees. The root (of a subtree) is created by repeatedly partitioning its $P$ pages into two sets, until there are $\floor{C_B}$ subsets, each containing $\ceil{P/C_B}$ pages, where $C_B$ is the branch capacity. For each partition, TGS considers $O(C_B)$ splits in every dimension to identify the split that minimizes a given cost function (e.g. sum of the volumes of the MBBs). This process is recursively applied to the root's child nodes to build the R-Tree. TGS is very expensive to build\cite{arge_priority_2004, qi_packing_2020, yang_platon_2023}, and has fluctuating query performance\cite{qi_packing_2020}.

OMT~\cite{lee_omt_2003} is a top-down variant of STR that starts with a root containing the entire dataset. The height of a branch node $n$ is computed as $h = \ceil{\log_{C_B}{P_{n}}}$, where $C_B$ is the branch capacity, and $P_{n}$ is the total number of pages contained in $n$. Since $n$ is at height $h$, all its child nodes would contain $P_\texttt{child} = C_B^{h-1}$ number of pages, implying that $n$ has $\ceil{P_n/P_\texttt{child}}$ child nodes. Similar to STR, the pages are sorted and tiled, so that each dimension has $\floor{|n|^{\frac{1}{d}}}$ tiles. The tiles that contain a singular page become leaf nodes, and the tiles with multiple pages undergo the packing procedure recursively.

Waffle~\cite{moti_waffle_2022} uses a bottom-up bulk loading procedure that involves two steps. A first step creates the leaf level by recursively sorting and splitting the data points on the longest dimension, until each resulting subspace contains a single page. The split point corresponds to the entry that is nearest to the median and at the boundary of one of the pages, i.e, the entry ranked $C_L \times \floor{\frac{\ceil{N / C_L}}{2}}$. To create the next level, a second step reuses splits created for the leaf level in the order of their creation. The splitting stops when no subspace exceeds $C_B$ leaf nodes. This continues recursively until at most $C_B$ nodes remain at the top, which form the child nodes of the root. Waffle, similar to STR and OMT, incurs zero overlap between nodes at the same level.

\Cref{fig:snapshots} displays the leaf nodes of various bulk loaded indexes for 3 million points\footnote{We only use 3 million points because the leaf nodes of the full dataset are too dense to visualize.} of the OSM dataset~\cite{osm} (1 billion points).
The disk page size is set to 4KiB, yielding a maximum leaf node capacity of $C_L=341$ points for all indexes. \Cref{tab:snapshot-stats} illustrates the number of leaf nodes, and the total area and perimeter of leaf nodes for each method on the full dataset. KDB-Trees do not emphasize on space utilization and involve a high count of leaf nodes. R-tree packing schemes and Waffle achieve the minimum number of leaf nodes because they are fully packed. The total area of Hilbert-Tree leaf nodes exceeds the data space because of node overlaps. STR, TGS, and OMT have low total area, but high perimeter due to elongated nodes. In TGS, as suggested by its authors, we use the split that minimizes the sum of the area of the resulting partitions. Notably, the splits only minimize the area for the immediate partitions, and does not necessarily result in minimum area for the whole index. Moreover, this produces elongated nodes with the highest perimeter, consistent with the findings reported in \cite{yang_platon_2023}. Waffle creates the partitions with the best characteristics, but as shown in the experimental evaluation, it is expensive to build.

\begin{table}[t]
	\begin{tabular}{l | l l l}
		\toprule
		Index    & Count   & Perimeter & Area   \\
		\cmidrule(lr){1-4}
		KDB-Tree & 4194304 & 0614769   & 064280 \\
		Hilbert  & 2932552 & 1983297   & 137157 \\
		STR      & 2932552 & 1470623   & 049308 \\
		TGS      & 2932552 & 100757580 & 046245 \\
		OMT      & 2932552 & 1235628   & 039789 \\
		Waffle   & 2932552 & 0436979   & 039389 \\
		\bottomrule
	\end{tabular}
	\caption{Total count, perimeter, and area of leaf nodes}
	\label{tab:snapshot-stats}
\end{table}

In recent years there is a large amount of work on \textit{learned} indexes that replace internal nodes with machine learning models (e.g., artificial neural networks), or utilize machine learning to enhance the index capabilities. Several multidimensional learned indexes (e.g., Flood~\cite{nathan_flood_2020}, Tsunami~\cite{ding_tsunami_2020}) assume a known data distribution and query workload, or involve approximate querying (e.g., RSMI~\cite{qi_rsmi_2020}). Moreover, they focus primarily on in-memory operations and lack in the spatial domain (e.g., not supporting $k$NN queries \cite{wang_zm_2019, nathan_flood_2020}).
Even disk-based learned indexes such as LISA~\cite{li_lisa_2020}, and PLATON~\cite{yang_platon_2023} are very slow to build since they involve model training.
For instance, PLATON uses Monte Carlo tree search to build models based on a given workload. While all the aforementioned conventional bulk loading methods require under an hour to bulk load the full OSM dataset, PLATON takes about 10 hours for 10\% of the dataset. Since our aim is efficient index building, we do not consider learned multidimensional indexes as competitors of the proposed methods.

\begin{figure*}[t]
	\centering
	\begin{subfigure}{0.33\textwidth}
		\includegraphics[width=0.98\textwidth]{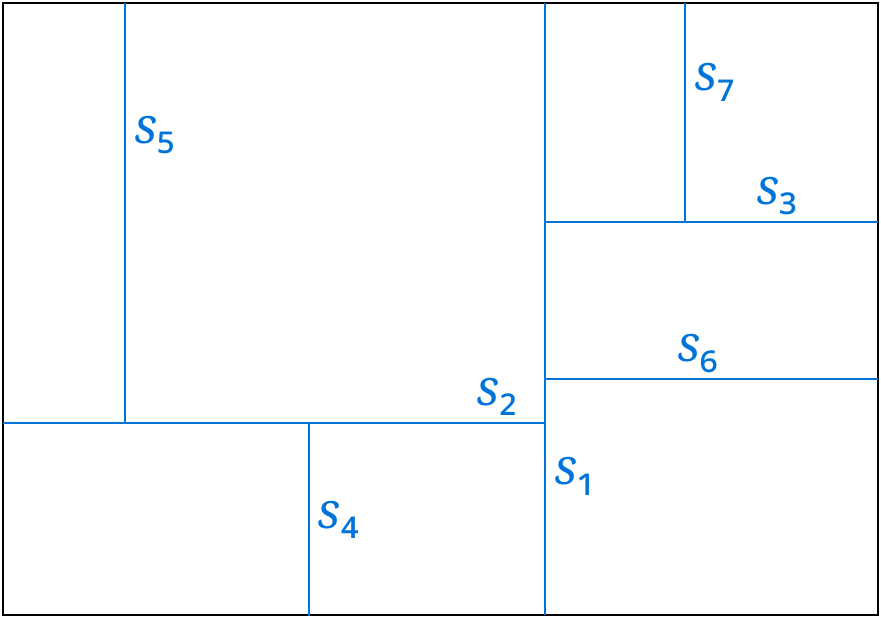}
		\caption{Partition Scheme}
		\label{fig:partition-scheme}
	\end{subfigure}
	\begin{subfigure}{0.33\textwidth}
		\includegraphics[width=0.98\textwidth]{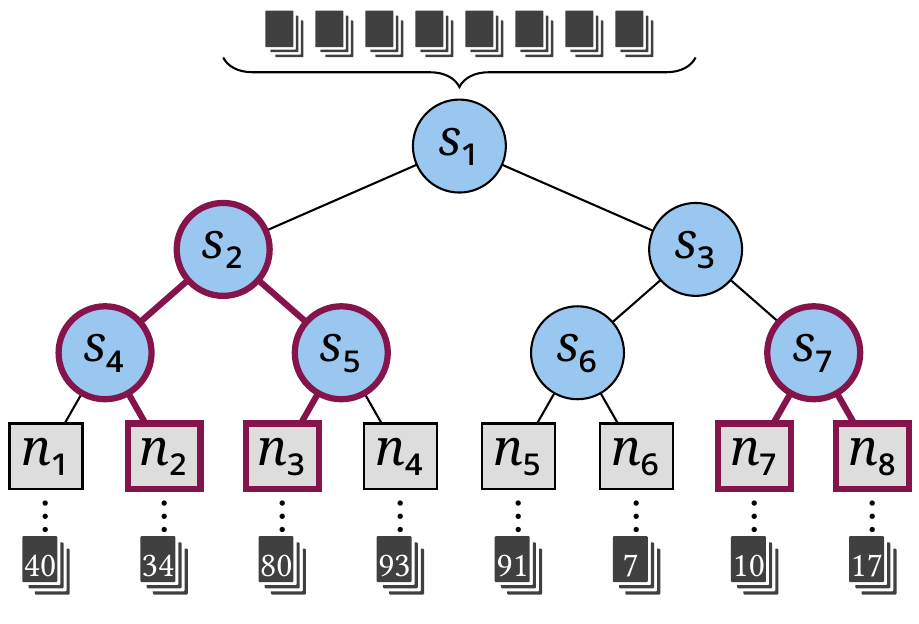}
		\caption{Major SplitTree $MST$}
		\label{fig:split-tree}
	\end{subfigure}
	\begin{subfigure}{0.33\textwidth}
		\includegraphics[width=0.98\textwidth]{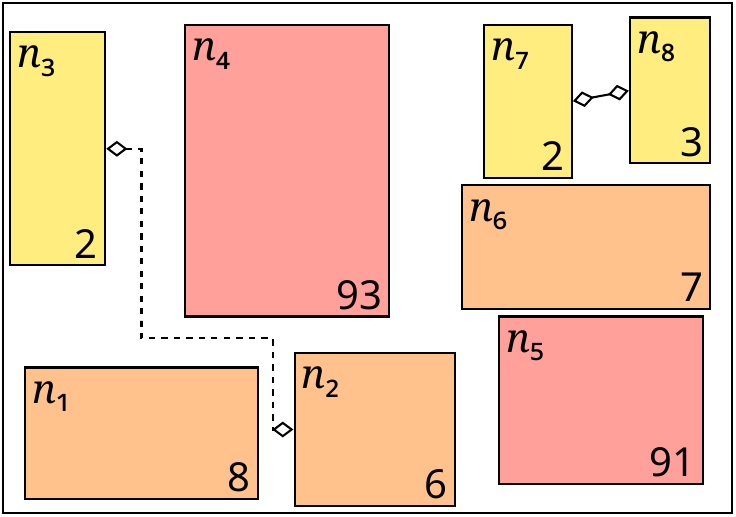}
		\caption{Root entries of FMBI}
		\label{fig:node-merging}
	\end{subfigure}
	\caption{FMBI Example}
	\label{fig:init}
\end{figure*}

\subsection{Other Related Work}\label{subsec:Adaptive Indexes and Database Cracking}
Bulk loading generates a complete and fixed index, independently of the query workload. On the other hand, \emph{adaptive indexes} are generated progressively as a response to query processing~\cite{karp_defer_1988}. Consequently, parts of the index that participate in more queries, are more refined than the rest. Recently the concept of adaptive indexes has received attention in \emph{database cracking} ~\cite{idreos_cracking_2007}.
QUASII ~\cite{pavlovic_quasii_2018} is an adaptive KDB-Tree based index that applies cracking to one dimension per tree level. At the top level, QUASII cracks along the query's extent on the first dimension generating a piece on that dimension. Then, at the next level it indexes the piece corresponding to the query's extent on the second dimension, and so on. After the query is processed, the index is a wide tree of $D$ levels, each associated with a single dimension.
AKD (adaptive KD-tree) ~\cite{nerone_adaptive_2021} is based on a similar idea, but it adds up to $2D$ new tree levels, one for each bound along each dimension.
Unlike QUASII and AKD, which index one dimension per level, the AIR-tree ~\cite{zardbani_adaptive_2023} maintains all dimensions at each level. The index is a main memory R-tree, generated incrementally, by splitting nodes that intersect incoming queries. Nodes stop being cracked, when their entries fall below a threshold, in which case the index reaches its steady state. \cite{jensen_revisiting_multi_ada_2021} contains an experimental evaluation of multidimensional adaptive indexes for main memory.

The term \emph{adaptive indexing} has also been used in different contexts. In ~\cite{tao_ais_2002}, it refers to indexes where the node size (i.e., number of disk pages per node) is based on the data and query characteristics (e.g., in parts of the data space where range queries have large extents, nodes may occupy multiple pages). Assuming that the query distribution is known in advance, ~\cite{achakeev_query_adaptive_2012}  bulk loads an R-tree optimized for the query workload. In both ~\cite{tao_ais_2002} and ~\cite{achakeev_query_adaptive_2012} the index is full, as opposed to progressively refined.

Spatial partitioning is the process of dividing an entire space into multiple disjoint subspaces that are usually managed by different servers. Each server receives and processes queries about data within its own subspace. A number of systems (~\cite{Eldawy2015,  SpatialHadoop, GeoSpark, Simba, Sphinx}) apply STR~\cite{leutenegger_str_1997} based techniques to perform space partitioning. R*-Grove ~\cite{R*-Grove} samples the data points and partitions the space using the R*-Tree split algorithm ~\cite{beckmann_rstar_1990}. SATO ~\cite{SATO}, also based on sampling, aims at minimizing the number of multidimensional objects crossing partitions. ~\cite{papadopoulos_parbulk_2003} studies the problem of parallel bulk loading for spatial data reside in distributed servers.


\section{Full Bulk Loading and FMBI}\label{sec:fmbi}
This section describes techniques for bulk loading the full index, called FMBI (Fast Multidimensional Bulkloaded Index). FMBI utilizes concepts of both data and space partitioning. Specifically, similar to R-trees, the node extents are minimum bounding boxes (MBBs) that tightly cover the underlying entries. On the other hand, node splits occur on the median of the longest dimension, similar to Spread KDB-Trees. Bulk loading is top-down and involves five concrete steps.

\subsection*{Step 1: Initial Partitioning}
We assume a main memory buffer of $M$ pages, such that $M > C_B$ where $C_B$ is the branch capacity. The bulk loading process starts by reading $\alpha \cdot C_B$ random pages of the dataset, where $\alpha = \floor{M/C_B}$, and sorting their points in-memory, on the longest dimension. The last point of the $\floor{(\alpha \cdot C_B)/2}^\text{th}$ \emph{sorted} page is the partition point, and its coordinate on the longest dimension, constitutes the split, which becomes the root of a \emph{Major} SplitTree, referred to as $MST$. Based on the split point, the first $\floor{(\alpha \cdot C_B)/2}$ sorted pages form the first subspace, while the rest form the second subspace\footnote{Using the above mechanism, all points of each sorted data page are assigned to the same subspace.}. Next, we recursively partition these subspaces on their longest dimension, until each subspace contains $\alpha$ pages. For every new split point, a pointer is stored in its parent at $MST$. If the split dimension changes (between a split point and its parent), sorting on the new dimension is required, but it is always performed in main memory. Once completed, $MST$ contains $C_B - 1$ splits that partition the space into $C_B$ subspaces, each with $\alpha$ full pages.

\Cref{fig:init} shows an example assuming branch node capacity $C_B = 8$. The first split $s_1$ occurs on the (longest) $x$- dimension, while its children $s_2$ and $s_3$ are split on the $y$-axis. At the end of Step 1, there is a total of 7 split points and 8 subspaces. In addition, we maintain the MBBs of the subspaces, which will form into FMBI root entries during the subsequent steps.

\subsection*{Step 2: Distribution of Remaining Pages}

This step distributes the points of the remaining pages (excluding the $\alpha \cdot C_B$ pages already scanned) into the subspaces of Step 1.
Initially, each subspace is deemed to be \emph{active}, and its $\alpha$ full pages are kept in main memory. At the first insertion in some subspace, we allocate it a new buffer page.
A search on $MST$ determines the subspace covering each data point $p$; $p$ is inserted in the corresponding buffer page, and the MBB of the subspace is adjusted if necessary.
If the buffer reaches its limit when allocating a new page to an active subspace, we flush all its full pages to the disk, which renders it \emph{inactive}. Each inactive subspace retains a single memory page, which is flushed when it fills.
When all points have been exhausted, Step 2 terminates. Continuing the running example, the number of pages after Step 2 is shown below each subspace in \Cref{fig:split-tree}.

\subsection*{Step 3: Refinement of Sparse Subspaces}

The goal of this step is to create the FMBI subtree for each subspace that can fit in the available buffer, referred to as \emph{sparse}. Active sparse subspaces are processed first because their pages are already in-memory. \Cref{algo:process-sparse} describes the recursive procedure, where the initial input is the list of pages of a subspace $n$, and the final output is the list of entries of the corresponding FMBI node. The procedure follows a post-order traversal of a \emph{minor} SplitTree for $n$ ($mST_n$), where at each $mST_n$ node visit, it processes a list of pages, $\mathcal{P}$. If $\abs{\mathcal{P}} = 1$, it returns a FMBI leaf node entry generated from that single page. Otherwise, it sorts the points of $\mathcal{P}$ on the longest dimension, and partitions them into two halves, $\mathcal{P}_1$ and $\mathcal{P}_2$, with $\floor{\abs{\mathcal{P}}/2}$ and $\ceil{\abs{\mathcal{P}}/2}$ pages, respectively. The two halves undergo the same process recursively, returning two sets of entries $ne_1$ and $ne_2$. If the sum of entries does not exceed $C_B$, the output is a concatenated list of $ne_1$ and $ne_2$. Otherwise, \Cref{algo:process-sparse} returns two FMBI branch node entries $nb_1$ and $nb_2$, generated from $ne_1$ and $ne_2$.

\begin{algorithm}
	\begin{algorithmic}[1]
		\Procedure{generate{\_}entries}{\emph{Page}[] $\mathcal{P}$}
		\If {$\abs{\mathcal{P}} = 1$}
		\State $nl \gets$ leaf node entry of $\mathcal{P}$
		\State \Return [$nl$] \Comment{List containing a single entry}
		\EndIf
		\State Sort points of $\mathcal{P}$ on longest dimension
		\State Partition $\mathcal{P}$ into $\mathcal{P}_1$ and $\mathcal{P}_2$
		\State \emph{NodeEntry}[] $ne_1 \gets$ \emph{generate{\_}entries}($\mathcal{P}_1$)
		\State \emph{NodeEntry}[] $ne_2 \gets$ \emph{generate{\_}entries}($\mathcal{P}_2$)
		\If {$\abs{ne_1} + \abs{ne_2} \leq C_B$}
		\State $ne \gets$ Concatenated list of $ne_1$ and $ne_2$
		\State \Return $ne$
		\Else
		\State $nb_1 \gets$ branch node entry of $ne_1$
		\State $nb_2 \gets$ branch node entry of $ne_2$
		\State \Return [$nb_1, nb_2$] \Comment{List containing $nb_1$ and $nb_2$}
		\EndIf
		\EndProcedure
	\end{algorithmic}
	\caption{Procedures for refining subspaces}
	\label{algo:process-sparse}
\end{algorithm}

\begin{figure}[t]
	\includegraphics[width=\linewidth]{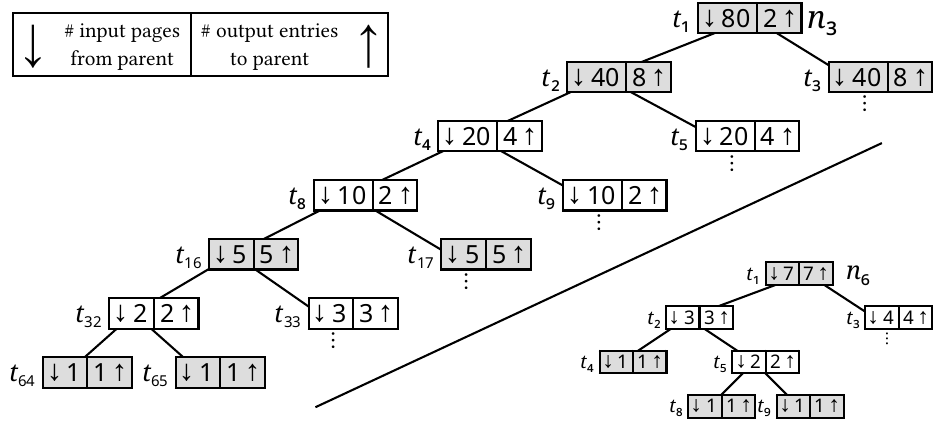}
	\caption{Minor SplitTrees of $n_3$ and $n_6$}
	\label{fig:minor-splittree}
\end{figure}

\Cref{fig:minor-splittree} illustrates the application of \Cref{algo:process-sparse} on $n_3$ containing 80 pages (see \Cref{fig:split-tree}), and the corresponding minor SplitTree $mST_3$. For each $mST_3$ node, the left cell shows the number of pages received as input, and the right cell the number of FMBI node entries returned.
The post-order traversal starts from the root and first creates two FMBI leaf entries $nl_{64}$, $nl_{65}$ that correspond to the two leftmost leaf nodes $t_{64}$ and $t_{65}$ of $mST_3$. When \Cref{algo:process-sparse} backtracks to their parent $t_{32}$, since $\abs{nl_{64}} + \abs{nl_{65}} = 2 < 8$, it returns the concatenation of $nl_{64}$ and $nl_{65}$ to $t_{16}$, which in turn outputs five concatenated entries ($nl_{64},..,nl_{69}$) to $t_8$. At $t_8$ the total number of input entries (10), received from $t_{16}$ and $t_{17}$, exceeds $C_B = 8$. This leads to the addition of two FMBI branch entries $nb_{16}$ and $nb_{17}$, corresponding to $mST_3$ nodes $t_{16}$ and $t_{17}$. The process continues all the way to the root of $mST_3$, where two FMBI root entries are created for $t_2$ and $t_3$. The grey cells indicate $mST_3$ nodes that generate FMBI entries. The bottom right corner of \Cref{fig:minor-splittree} illustrates a simpler case for $n_6$, whose subspace is divided into 7 FMBI leaf entries, each corresponding to a leaf node of $mST_6$.

When an active subspace is finalized, its buffer pages and $mST$ are released from main memory. After all active subspaces are processed, each inactive sparse subspace is reloaded and refined in the same way. \Cref{fig:node-merging} indicates for each sparse subspace, its MBB and the number of root entries of their FMBI subtree after refinement. Subspaces $n_4$ and $n_5$ are \emph{dense}, i.e., their size exceeds the available buffer, and will be processed at Step 5.

\subsection*{Step 4: Merging of Underflowed Branches}

Branch nodes with no more than $C_B/2$ entries (e.g., $n_3, n_7, n_8$ in \Cref{fig:node-merging}) are \emph{underflowed}. To mitigate their negative effect on query performance, we invoke a bottom-up strategy that merges them \emph{conceptually}.
Concretely, we perform a post-order traversal of the Major SplitTree, where for each underflowed branch $n_u$, we find the subspace $n_v$ with the lowest common ancestor to $n_u$ at the $MST$, such that the total number of their entries is within $C_B$. The aggregated contents of both nodes are stored on the same disk page, but the root of FMBI maintains two separate entries for $n_u$ and $n_v$. This process does not introduce additional I/O during query processing since merged pages are only accessed if some of their nodes potentially contain query results.

\Cref{algo:merge-branch} describes the pseudocode for merging. The input is a pointer $ptr$, which initially points to the root of $MST$. A recursive call terminates when $ptr$ points to a subspace. If the corresponding node $n$ is processed (i.e., sparse), the call returns $n$ as a candidate for merging. Otherwise, if $n$ is dense, it returns $\phi$. When $ptr$ points to a split $s$, the recursive calls to its left and right subtrees return $n_l$ and $n_r$. If one of $n_l$ or $n_r$ is $\phi$, the other node is returned without attempting to merge (this also covers the case where both $n_l$ and $n_r$ are $\phi$). If the total number of entries of $n_l$ and $n_r$ is within $C_B$, $n_l$ and $n_r$ are merged. Otherwise (no merge is possible) the algorithm returns the node with the fewer number of entries, as a potential candidate for merging upstream. After the algorithm terminates, there is at most one underflowed branch remaining and $MST$ is deleted.

\begin{algorithm}[t]
	\begin{algorithmic}[1]
		\Procedure{merge{\_}branches}{\emph{Pointer} $ptr$}
		\If {$ptr$ points to a leaf of MST}
		\State $n \gets$ MST node pointed by $ptr$
		\If {$n$ is processed}~\Return $n$
		\Else~\Return $\phi$ \Comment{$n$ is dense (not processed)}
		\EndIf
		\EndIf
		\State $s \gets$ split pointed by $ptr$
		\State $n_l \gets$ \emph{merge{\_}branches}(left pointer of $s$)
		\State $n_r \gets$ \emph{merge{\_}branches}(right pointer of $s$)
		\If {$n_l$ is $\phi$}~\Return $n_r$
		\EndIf
		\If {$n_r$ is $\phi$}~\Return $n_l$
		\EndIf
		\If {total number of entries in $n_l$ and $n_r$ is within $C_B$}
		\State \Return \emph{merge}($n_l, n_r$)
		\Else \If {$n_l$ has fewer entries than $n_r$}
		\Return $n_l$
		\Else~\Return $n_r$
		\EndIf
		\EndIf
		\EndProcedure
	\end{algorithmic}
	\caption{Procedure for merging the branches}
	\label{algo:merge-branch}
\end{algorithm}

In the example of \Cref{fig:init}, the initial input of \Cref{algo:merge-branch} is the root  $s_1$ of $MST$. The processing of $s_4$ returns $n_2$ since it has fewer entries (6) than its sibling $n_1$ (8). Similarly, $s_5$ returns $n_3$, as $n_4$ is dense and unprocessed. After receiving $n_2$ and $n_3$, $s_2$ merges them, and their entries (6 and 2, respectively) are stored in the same disk page. In the right subtree, $n_7$ and $n_8$ are merged at $s_7$ with a total of 5 entries in their shared disk page.

\subsection*{Step 5: Processing of Dense Subspaces}
Since dense subspaces (e.g., $n_4$, $n_5$) contain more data pages than the available buffer, they cannot be refined using Steps 3 and 4. Instead, each dense subspace $n$ is treated as new dataset to be bulk loaded. Accordingly, we apply all steps to the data points of $n$ and generate an index $FMBI_n$. The root of $FMBI_n$ becomes an entry in the root of FMBI.

\begin{figure}[t]
	\centering
	\begin{subfigure}{0.49\textwidth}
		\includegraphics[width=\linewidth]{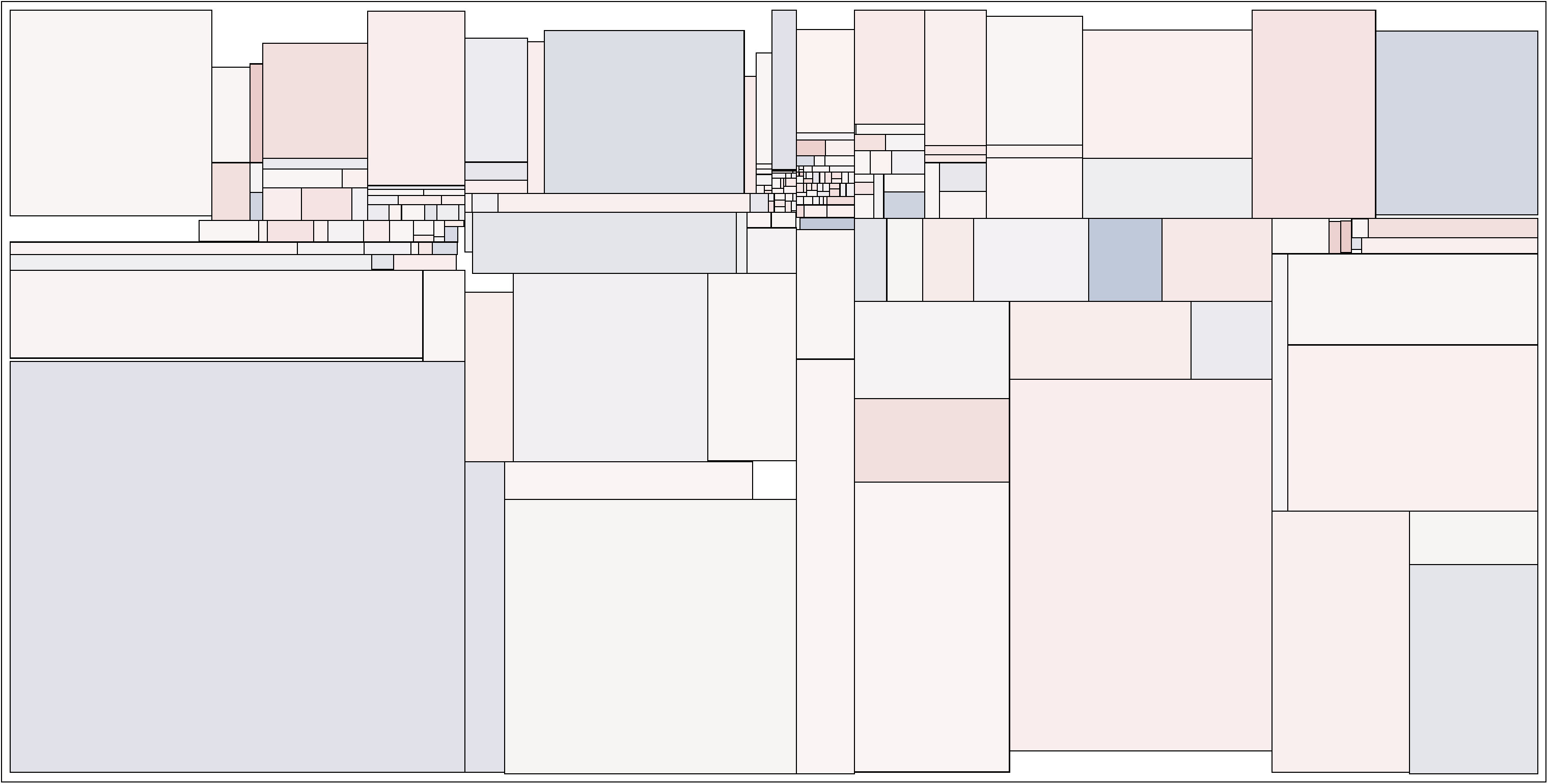}
		\caption{Root entries of full dataset}
		\label{fig:fmbi-root-nodes}
	\end{subfigure}
	\begin{subfigure}{0.49\textwidth}
		\includegraphics[width=\linewidth]{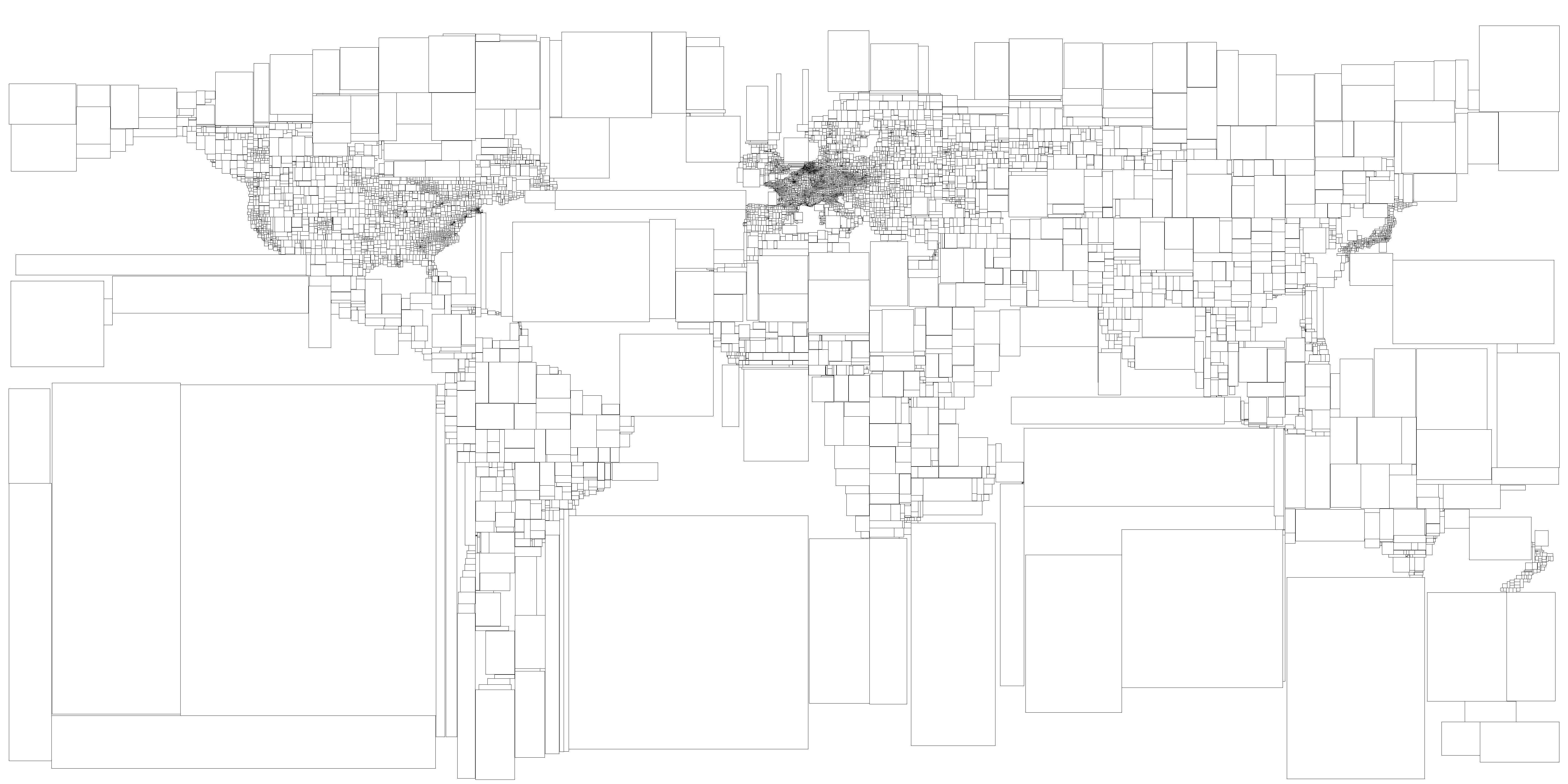}
		\caption{Leaf nodes}
		\label{fig:fmbi-osm}
	\end{subfigure}
	\caption{FMBI bulk loaded with OSM}
	\label{fig:fmbi}
\end{figure}

\Cref{fig:fmbi-root-nodes} shows the root entries of the final FMBI, after bulk loading OSM (the full dataset). The color of each subspace indicates the number of entries in its subtree, where red (grey) subspaces have more (fewer) points than average. Nevertheless, the variance is rather small, as the cardinality of the subspace with the maximum (minimum) number of entries is $1.06$ ($0.92$) times that of the average, suggesting that FMBI is rather balanced.
\Cref{fig:fmbi-osm} displays the leaf nodes of FMBI under the same settings as in \Cref{fig:snapshots}. When bulk loaded with the full OSM dataset, FMBI has 2932651 leaves, with total perimeter 432743, and area 39310. Comparing with \Cref{tab:snapshot-stats}, FMBI has marginally more leaves than the fully packed indexes, while exhibiting the lowest total area and perimeter.

\section{Adaptive Bulk Loading and AMBI}\label{sec:ambi}

Similar to the other multidimensional indexes, the proposed techniques apply top-down traversal for query processing. A query starts from the root, and recursively visits each node that may contain results. The search terminates at leaf nodes, where the corresponding pages are scanned and filtered to aggregate all qualifying points. Whereas conventional indexes, including FMBI, are built in advance (i.e., before the first query), in AMBI (Adaptive Multidimensional Bulkloaded Index) the index is built on-demand, when unprocessed nodes are encountered during query processing. We assume that the query type or distribution is not known in advance.

\subsection{Query Based Index Refinement}
In the absence of an index, each query necessitates a sequential scan of the data file. When the first query is received, AMBI initializes the root with Step 1, i.e., by loading $\alpha \cdot C_B$ pages in-memory and building the major SplitTree $MST$.
Similar to FMBI, at Step 2 all subspaces of $MST$ are initially \emph{active}, and their pages are kept in-memory, while the rest of the dataset is distributed. However, instead of deactivating the subspace when the buffer reaches its limit, AMBI maintains a max-heap $H$ of active subspaces based on their Euclidean distance from the query, and flushes the top of $H$. Consequently, subspaces that are \emph{unqualified} (i.e., those not containing query results) are deactivated first, whereas those that are \emph{qualified} are kept in main memory.

Eventually, even qualified subspaces may have to be deactivated, but more strategically.
If a qualified subspace $n$ contains $P_n<C_B$ pages, it is deactivated in the same way as an unqualified one. Otherwise, when $P_n \geq C_B$, AMBI generates its minor SplitTree $mST_n$ using $\beta \cdot C_B$ pages, where $\beta = \floor{P_n/C_B}$. All $mST_n$ subspaces are added to $H$ based on their distance from the query, so that they too become candidates for deactivation\footnote{The subspaces of $mST_n$ are less likely to to be useful and more likely to contain fewer than $C_B$ pages, which makes them suitable for deactivation.}. Any remaining pages of $n$ are distributed to $mST_n$'s subspaces. This process may be repeated recursively until the buffer is freed.

\begin{figure}[t]
	\includegraphics[width=\linewidth]{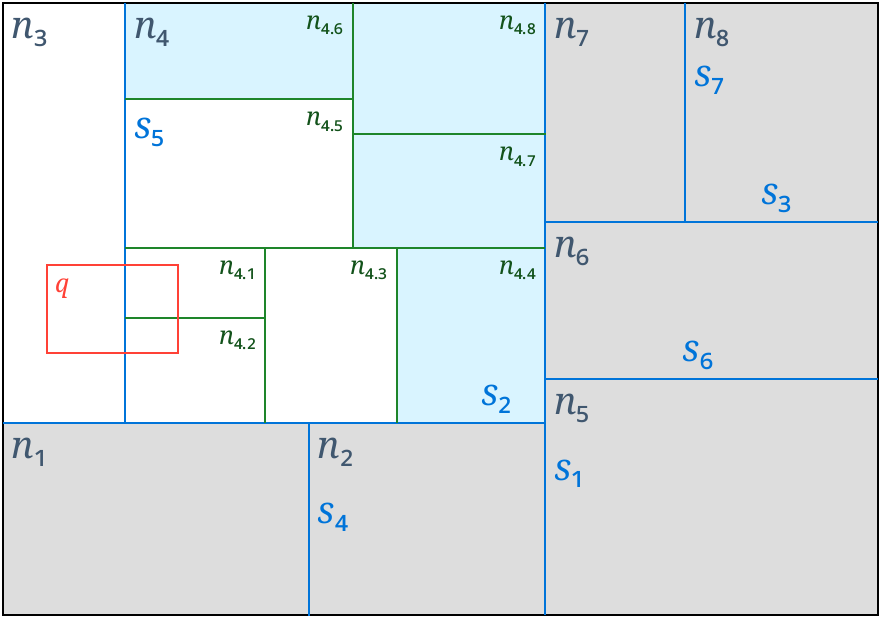}
	\caption{Adaptive subspace partitions}
	\label{fig:ada-subspaces}
\end{figure}

\Cref{fig:ada-subspaces} shows the partitions of \Cref{fig:init} in the context of adaptive indexing for a window query $q$ (red rectangle). Following Step 1, the space is partitioned into $C_B=8$ subspaces $n_1$ to $n_8$, which are inserted to $H$. As the buffer gets full during Step 2, subspaces are deactivated (highlighted in grey) in the reverse order of their distance from $q$. If the buffer fills when $H$ only contains qualified subspaces $n_3$ and $n_4$, $n_4$ is partitioned into $n_{4.1}$ to $n_{4.8}$ that replace $n_4$ in $H$. Some of those subspaces (shown in blue) are also subsequently deactivated. When distribution terminates the active subspaces are $n_3$, $n_{4.1}$, $n_{4.2}$, $n_{4.3}$ and $n_{4.5}$.

All active subspaces, including unqualified $n_{4.3}$, $n_{4.5}$, are refined by \Cref{algo:process-sparse} as they do not incur additional I/O cost. On the other hand, AMBI does not refine any inactive subspace $n_u$, whether sparse or dense. Instead, $n_u$ will be processed when it qualifies for some query in the future, using a minor SplitTree (if $n_u$ is sparse) or a major SplitTree (if $n_u$ is dense). Observe that an inactive subspace $n_u$, with $P_{n_u} < C_B$ pages will always generate $P_{n_u}$ leaf entries when processed. Thus, we can safely merge an unrefined subspace with a processed one, provided that their total number of entries is within $C_B$. This enables us to include all subspaces, irrespective of their constitution, in the merging process of \Cref{algo:merge-branch}. Naturally, nested subspaces are merged first, so that finalized entries of useful subspaces can participate in the merging at the root of $MST$.

\begin{figure}
	\begin{subfigure}{0.49\textwidth}
		\includegraphics[width=0.99\textwidth]{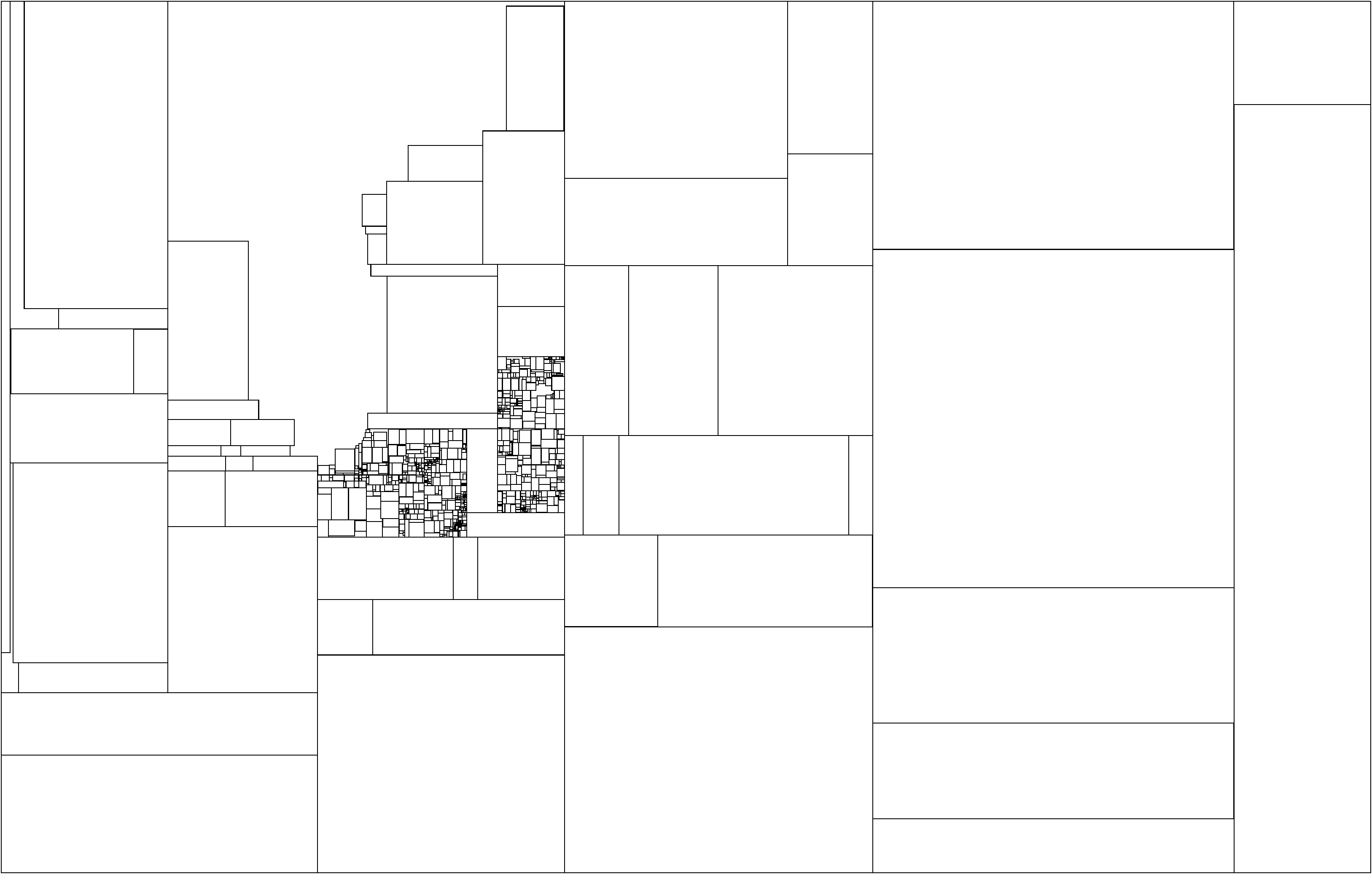}
		\caption{AMBI after 10 window queries focused on Germany}
		\label{fig:ambi-germany-10}
	\end{subfigure}
	\begin{subfigure}{0.49\textwidth}
		\includegraphics[width=0.99\textwidth]{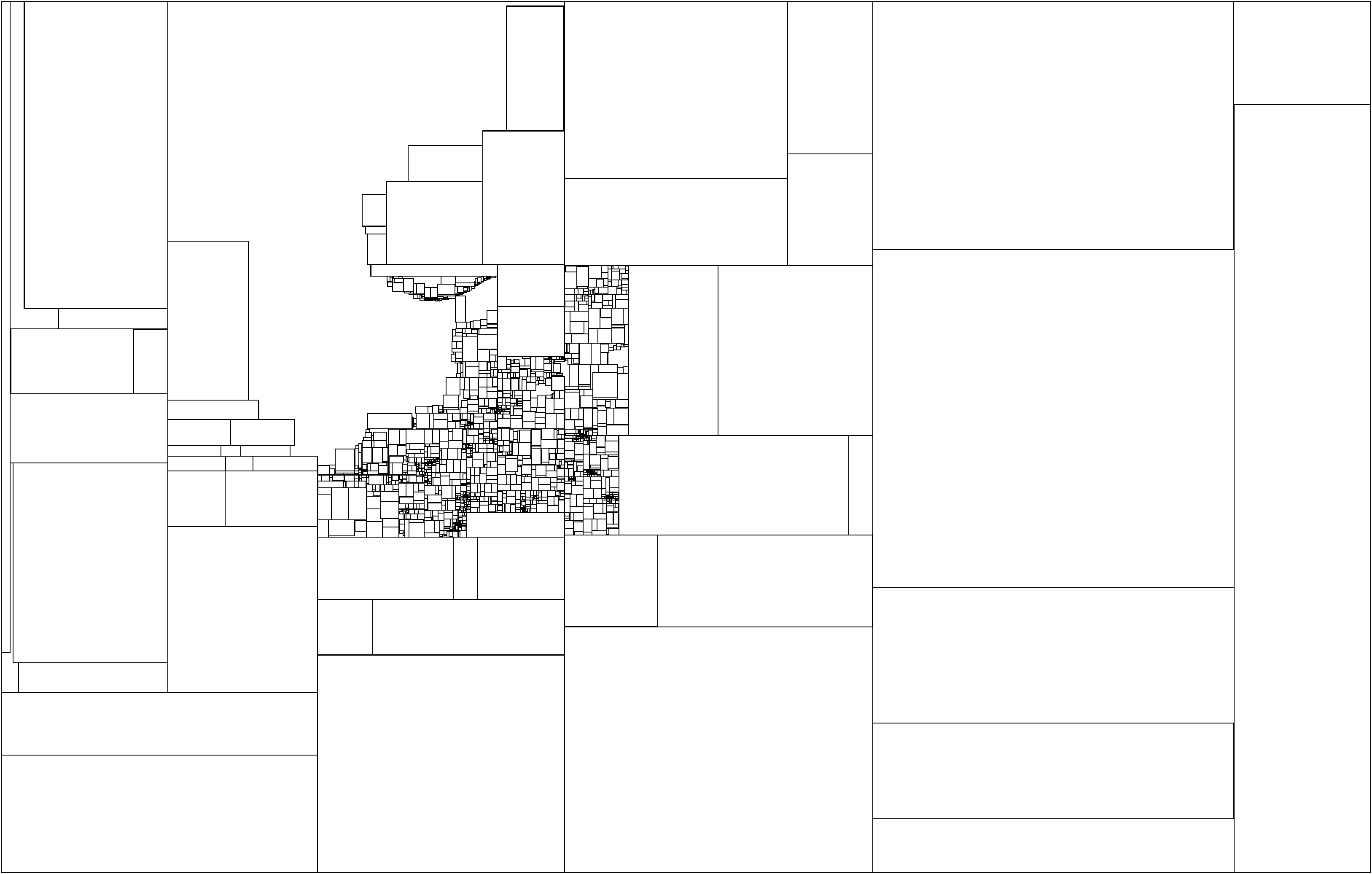}
		\caption{AMBI after 100 window queries focused on Germany}
		\label{fig:ambi-germany-100}
	\end{subfigure}
	\begin{subfigure}{0.49\textwidth}
		\includegraphics[width=0.99\textwidth]{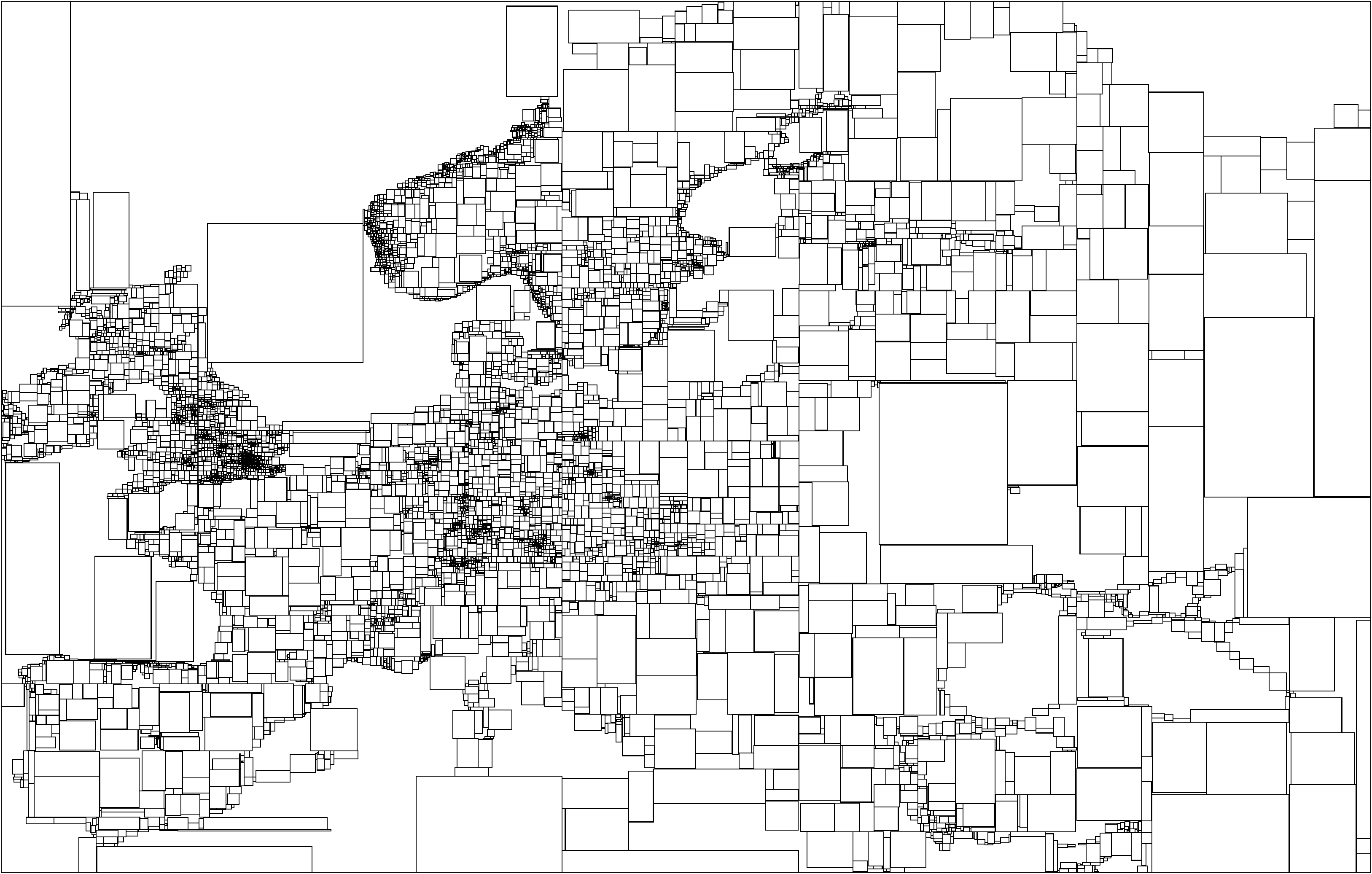}
		\caption{FMBI leaf nodes}
		\label{fig:fmbi-germany}
	\end{subfigure}
	\caption{Leaf level of Europe (OSM)}
	\label{fig:osm-germany}
\end{figure}

\Cref{fig:ambi-germany-10,fig:ambi-germany-100} illustrate the partial index generated by AMBI after 10 and 100 window queries focused on Germany, one of the densest areas in Europe. Only nodes containing data points in or around Germany are fully refined into leaf nodes. The rest of MBBs correspond to higher level nodes that remain unprocessed.
It is important to note that unlike other adaptive indexes (e.g., AIR~\cite{zardbani_adaptive_2023}), where the final tree depends on the order of queries, the set of AMBI nodes is independent of the query order. If the queries cover the entire data space, AMBI becomes identical to FMBI as shown in \Cref{fig:fmbi-germany}.

\subsection{Dynamic Updates}
Bulk loaded indexes can be maintained in the presence of updates using the algorithms of the corresponding dynamic structures. For instance R-trees generated by Hilbert, STR or OMT could utilize the insertion and deletion algorithms of R*-Trees ~\cite{beckmann_rstar_1990}, or other variants ~\cite{guttman_rtrees_1984, beckmann_revised_2009}. However, these algorithms can be expensive (e.g., R*-Trees re-insert all entries of an underflowed node) and unsuitable for update intensive workloads. To alleviate this problem, Waffle ~\cite{moti_waffle_2022} determines the update frequency of nodes based on their ratio of queries over updates, using two parameters \emph{fat} and \emph{tolerance}, so that only frequently queried nodes are continuously updated.

The proposed adaptive techniques provide a natural way to handle workload-based dynamic updates. Specifically, when a new point causes a leaf node to overflow, a new page is allocated to the node to accommodate that point as well as future insertions. Eventually, the node will be processed and refined, when it is accessed by some query. Similarly, underflows are only handled when the node is queried. Unlike Waffle, this lazy approach does not require the maintenance of statistical information per node (the counter of queries and updates), or the introduction of additional parameters (\emph{fat} and \emph{tolerance}). Compared to existing algorithms for dynamic multidimensional indexes, this is expected to drop the cost significantly for update intensive workloads, especially when updates arrive in bursts.

\section{Parallel Bulk Loading and Distributed Query Processing}\label{sec:parallel}
The scan-based nature of the proposed techniques enables their effective extension to distributed systems and spatial partitioning. We assume a central server that receives the data points and queries. The server distributes the data points to $m$ local servers, each responsible for a region of the data space. In this setting FMBI and AMBI can be bulk loaded in parallel, utilizing the resources (e.g., disk, buffer) of all servers. We first discuss the adaptation of FMBI.

Assuming a main memory buffer of $M$ pages, the central server partitions $\gamma \cdot m$ random pages of the dataset, where $\gamma = \floor{M/m}$, to $m$ subspaces, using a SplitTree with $m-1$ splits. The $\gamma$ pages of each subspace $n_i$ are sent to the corresponding local server $l_i$. Subsequently, every $l_i$ generates a local $FMBI^i$, indexing $n_i$. Specifically, the central server distributes the points of the remaining ($P - \gamma \cdot m$) pages to the local servers. Assume that the available buffer at server $l_i$ is $M_i$. When $l_i$ receives $\delta \cdot C_B$ pages of points within $n_i$, where $\delta = \floor{M_i/C_B}$, it performs its own Step 1, to partition $n_i$ into $C_B$ subspaces. Steps 2 to 5 form these subspaces into the root entries of $FMBI^i$, as discussed in \Cref{sec:fmbi}.

The cost incurred at the global server is equal to the total number $P$ of pages, since it reads the entire dataset once to compute the partitions and distribute the remaining points.
Let $P_i$ be the number of data pages for subspace $n_i$. Each local server $l_i$ creates $FMBI^i$ levels from top to bottom. Assuming that all subspaces are inactive, at each level, it iterates through all the $P_i$ pages, and generates $C_B$ partitions for the next level. After $L=\log_{C_B}(P_i/M_i)$ levels of partitions, the subspaces are partitioned in the main memory. Thus, the cost for each $l_i$ is:
$\sum_{l=0}^{L}{C_B^l \frac{P_i}{C_B^l}} = P_i \log_{C_B}(P_i/M_i)$. The parallel running time is determined by the local server with the highest cost \cite{beame_pqp_2013} ~\cite{qi_packing_2020}. If all servers have identical resources, this would be the $l_i$ with the largest number $P_i$ of data pages.

When the central server receives a query, it directs it only to \emph{qualified} local servers, i.e., those that may contribute results. For window queries, qualified servers are those whose subspace intersects the window. However, for nearest neighbor ($k$-NN) queries there is no predefined range. To overcome this problem, SpatialHadoop ~\cite{eldawy_spatial_hadoop_2015} processes a $k$-NN query $q$ in two rounds. Round 1 finds $k$ candidate NNs in the server covering $q$, and computes the distance $dist$ between $q$ and the $k$-th candidate. Round 2 searches for additional candidates in the local servers whose subspace intersects the circle centered at $q$ with radius $dist$. AQWA ~\cite{aly_aqwa_2015} processes the query in a single round, using histograms to identify qualified servers. Either method is compatible with the proposed techniques.

The extension of the above to adaptive indexing is straightforward. When the central server receives the first query, it applies Step 1 to build the partition scheme and determine the subspace of each local server. Then, at Step 2 it distributes to every local server the data points within its subspace. Each local server performs Steps 1 to 4, to build a partial AMBI, without incurring additional I/O\footnote{Operations that incur additional I/O, i.e., refinement of inactive (at Step 3) and dense (at Step 5) subspaces are deferred for later.}. The first query is processed by the central server that has to scan the entire data set anyway. When each new query arrives, it is directed to the qualified servers, which use the query to refine qualified nodes at their local index as discussed in Section \Cref{sec:ambi}.

\section{Experimental Evaluation}\label{sec:exp}
We compare the proposed FMBI, and its adaptive version AMBI, against the methods discussed in \Cref{sec:related-work}.
We exclude TGS because they are at least an order of magnitude more expensive at bulk loading and query processing than others (as shown in \Cref{tab:snapshot-stats}, most of their nodes are elongated). Similarly, learned indexes are excluded because of their large training time, and pre-requisite workload requirement.
For KDB-Trees we include the more efficient \emph{spread} variant ~\cite{friedman_kdbspread_1977}, bulk loaded by the algorithm of ~\cite{moti_waffle_2022}. To ensure fairness, the indexes were implemented within the same disk-based framework that we developed in Rust\footnotemark{}.
All experiments were conducted on an AMD Ryzen Threadripper 3960X 3.8GHz CPU with 64GiB RAM and 64-bit Ubuntu Linux operating system, with a disk-page size of 4KiB. We use the following real data sets:
\begin{enumerate}
	\item OSM~\cite{osm}: 1 billion 2D geolocations across the globe.
	\item NYCYT~\cite{nycyt}: 100 million 5D trip records of New York City yellow taxis in the year 2014. The dimensions are the $x$, $y$ coordinates of the pick up and drop off points, and the time.
	\item Additional experiments with uniform, guassian, and skewed data are included in the code repository\footnotemark[\value{footnote}].
\end{enumerate}
\footnotetext{The code will be open sourced post acceptance.}

\begin{figure*}[t]
	\centering
	\includegraphics[width=0.98\textwidth]{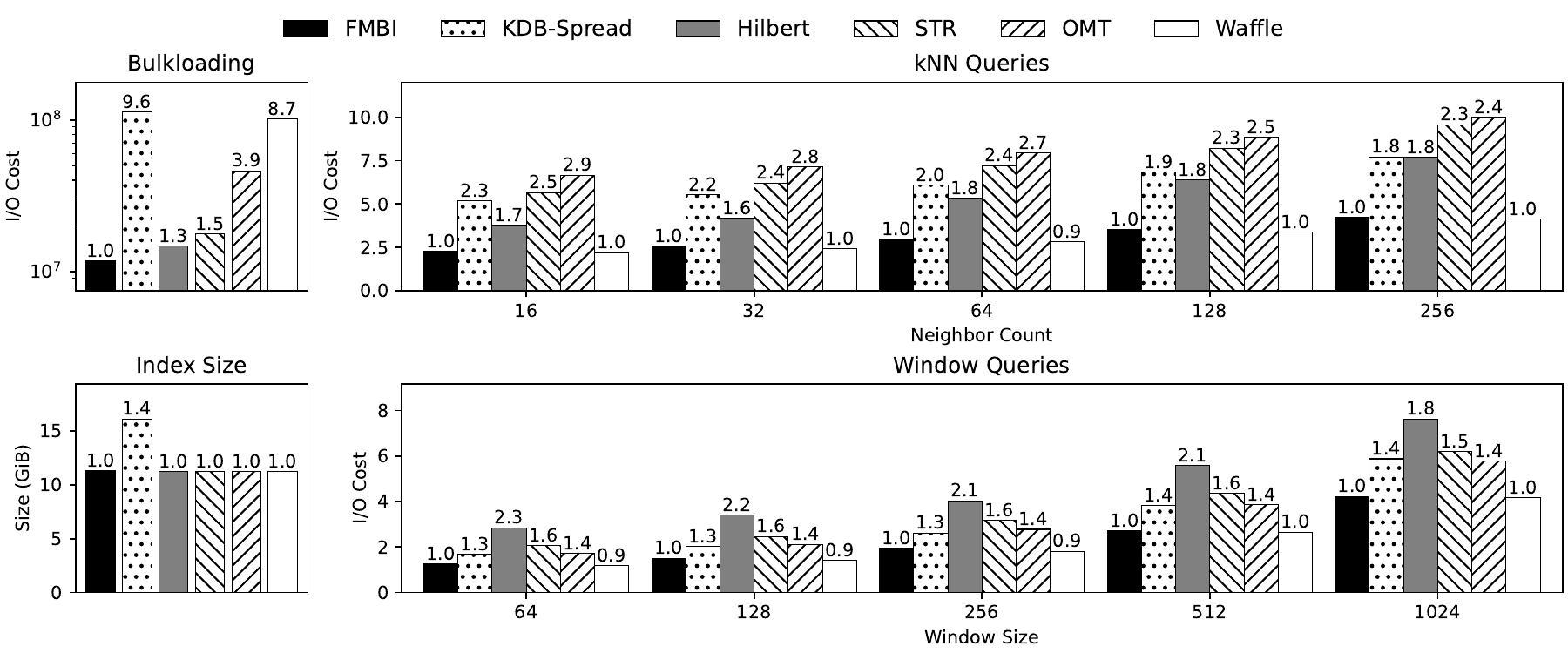}
	\caption{OSM | Non-adaptive}
	\label{fig:OSM-NA}
\end{figure*}

The first experiment focuses on the conventional (i.e., non-adaptive) bulk loading of OSM using an LRU buffer equal to 1\% of the dataset. Given the page size of 4KiB, the maximum capacity is $C_L=341$ for leaf nodes, and $C_B=204$ for branches, in all indexes (branch entries store bounding boxes and require two points per entry, instead of one for leaves). OSM occupies 2932552 disk pages, i.e., the same as the number of leaf nodes of fully packed indexes in \Cref{tab:snapshot-stats}. The 1\% buffer size corresponds to 29325 pages, and the value of $\alpha$ in FMBI is $143 = \floor{29325/204}$, i.e., Step 1 partitions into 204 subspaces, as shown in \Cref{fig:fmbi-root-nodes}, each containing 143 full pages. In order to study the effect of sampling in FMBI, we executed the experiment using 100 different samples (each with size 1\% of the dataset).  The average bulk loading cost in terms of page I/O, i.e., total number of page reads and writes was 11733245, and the maximum (minimum) was 11757239 (11727645). The difference in term of query processing cost are similarly negligible. The reported results correspond to the average values.

The top-left diagram in \Cref{fig:OSM-NA} illustrates the cost of building the full index. The numbers on top of each method indicate its relative performance compared to FMBI. Top-down approaches are the most expensive because they involve numerous applications of external sorting. Specifically, KDB-Trees are the slowest because of the larger number of leaf nodes (see \Cref{tab:snapshot-stats}), followed by Waffle and OMT. For bottom-up approaches, Hilbert outperforms STR because given the available buffer, it only performs external sorting once for the leaf level nodes. FMBI is naturally even faster as it avoids external sorting altogether. Moreover, as shown in the bottom-left diagram, FMBI has the same size as the indexes that pack leaf nodes fully, demonstrating the effectiveness of the merging process at Step 4.

\begin{figure*} [t]
	\centering
	\includegraphics[width=0.98\textwidth]{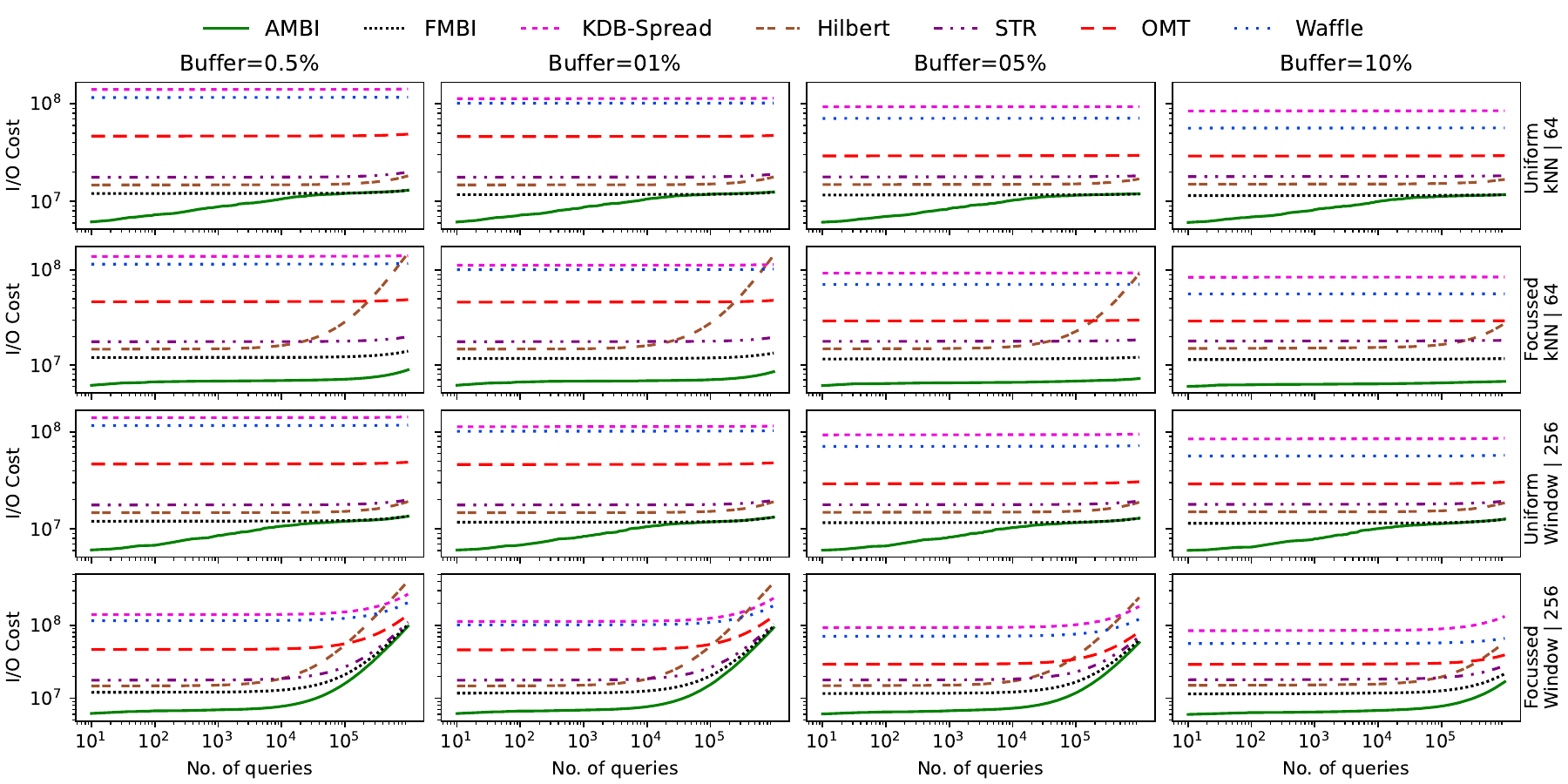}
	\caption{OSM | Adaptive}
	\label{fig:OSM-A}
\end{figure*}

The second column of diagrams in \Cref{fig:OSM-NA} measures the cost of $k$-nearest neighbor and window queries, as a function of $k$ and window size, given the 1\% LRU buffer.  For $k$-NN queries, $k$ ranges from 16 to 256. Each window query is a rectangle, whose aspect ratio is the same as the data space, and area is a percentage that ranges between $64/N$ and $1024/N$, where $N$ is the data cardinality. Every reported value is the average of 1000 queries, uniformly spread in the entire data space. As expected, the absolute number of disk pages retrieved increases with the output size for all methods. FMBI is almost as fast as Waffle, which, however, is $8.7$ times slower on index building. All other methods are consistently slower for all settings, which can be explained by the inferior node properties demonstrated in \Cref{tab:snapshot-stats}. The relative performance of these methods depends on the query characteristics.

The next experiment focuses on the adaptive bulk loading of OSM and the performance of AMBI. Each row of plots in \Cref{fig:OSM-A} corresponds to a query type, and every column to an LRU buffer size that ranges between 0.5\% to 10\% of the dataset. The diagrams in the first row illustrate the combined I/O cost of index building and 64-NN uniform queries over the entire space. Specifically, for non-adaptive methods including FMBI, we build the complete index in a single step, perform up to $10^6$ 64-NN, and plot the total (index building and cumulative query) cost as a function of the number of queries performed. For AMBI, the index is gradually refined as each query is processed. For all buffer sizes, AMBI is the fastest method, usually by orders of magnitude. As expected, the benefits of AMBI are more pronounced for a small number of queries.

According to \Cref{fig:OSM-NA}, for non-adaptive methods index building is 6-8 orders of magnitude more expensive than single queries, and, thus, it dominates the total cost. Therefore, there is no visible difference in their performance in \Cref{fig:OSM-A} as the number of queries reaches $10^6$. FMBI is the most efficient, followed by Hilbert, STR and OMT. Although Waffle has fast query processing, it is the second most expensive overall (after KDB-Trees) due to its high building cost. A large buffer size improves all methods because it leads to lower building cost (the speed of external sorting drops with larger buffer sizes) and better query performance (more index nodes remain in-memory).

The second row of \Cref{fig:OSM-A} repeats the experiment of the first row, but now all query points are \emph{focused} within the bounding box of Germany, which constitutes a dense area. This has a minimal effect on non-adaptive methods, whose cost is dominated by index building. On the other hand, for AMBI this also affects index building since the partial index only covers space potentially containing query results. Accordingly, AMBI does not converge to FMBI as the number of queries increases, and most of the data space (excluding Germany) remains non indexed (see \Cref{fig:osm-germany}).

Rows 3 and 4 of \Cref{fig:OSM-A} illustrate the corresponding diagrams for uniform and focused window queries, which are similar to those for $k$-NN. The main difference is in the last row where the cost of query processing starts becoming visible for FMBI, Hilbert and STR after the first $10^4$ queries. This happens because focused window queries in a dense area have large output, and are more expensive than uniform queries of the same size.

\begin{figure*}[t]
	\centering
	\includegraphics[width=0.98\textwidth]{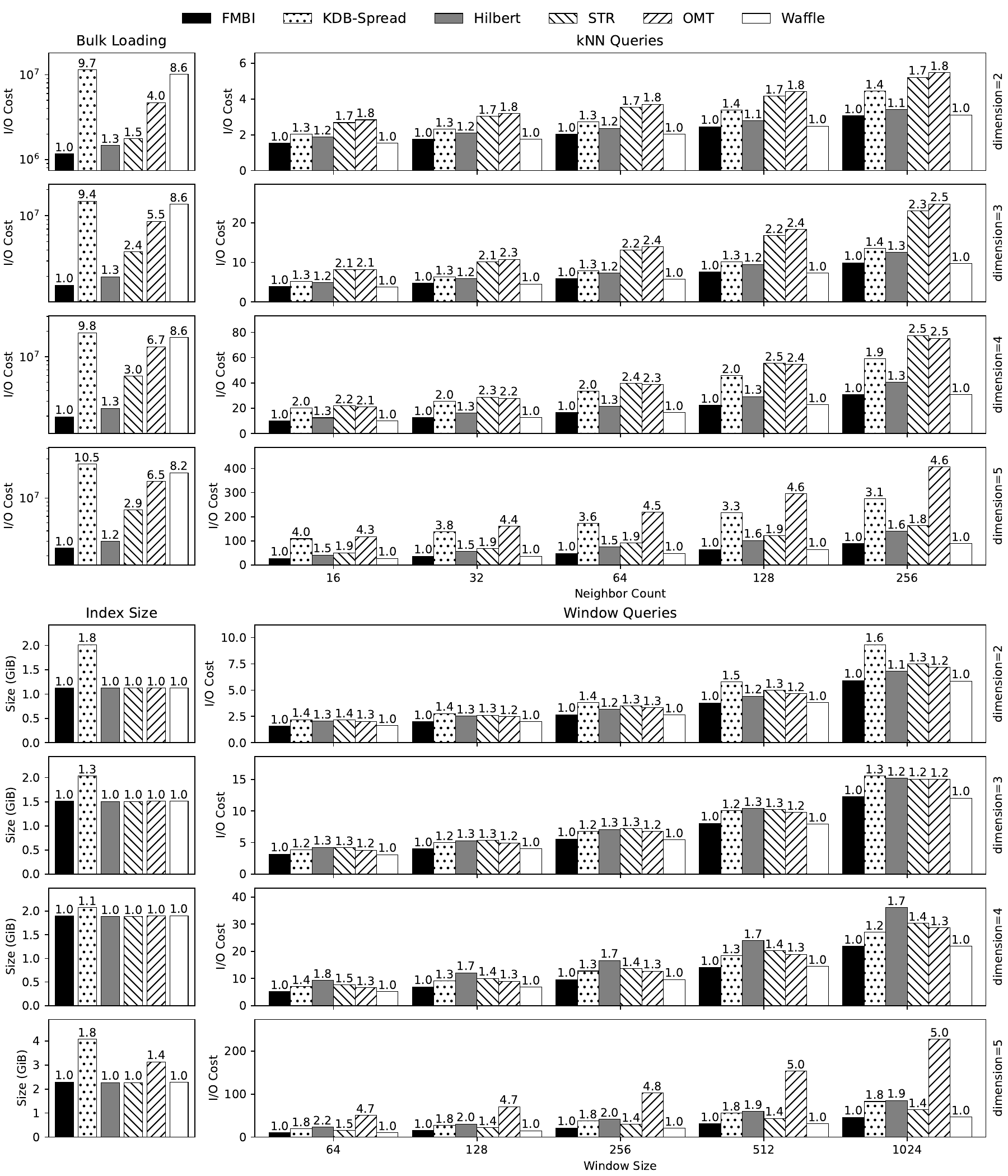}
	\caption{NYCYT | Non-adaptive}
	\label{fig:NYC-NA}
\end{figure*}

\Cref{fig:NYC-NA} evaluates the performance of non-adaptive indexes versus the number $d$ of dimensions using NYCYT. Each row of plots corresponds to a different value of $d$ between 2 and 5. When $d<5$, we select the first $d$ dimensions of the dataset. As shown in the diagrams of the left column, FMBI is the fastest to build for all values of $d$, while having the same size as the fully packed indexes \footnote{For $d=5$, the bulk loading procedure of OMT cannot pack nodes to their full capacity, which affects both the index size and the query performance.}. Regarding query processing, all queries are uniformly distributed in the data space. The absolute cost of all methods increases with $d$, as the tree fan-out decreases due to the additional dimensions. The effect is more evident on $k$-NN queries due to the dimensionality curse \cite{beyer_useful_nn_1999}, i.e., the volume of space that must be searched for candidate neighbors grows exponentially with $d$. Similar to the experiments for OSM, FMBI and Waffle are the most efficient for every query setting (but Waffle is 8.2 to 8.6 times slower to build). Compared to \Cref{fig:OSM-NA} the difference of the other methods relative to FMBI is lower because NYCYT is less skewed (e.g., it does not contain empty areas such as the oceans of OSM). However, the relative difference gradually increases with $d$, indicating that FMBI (and Waffle) scales better with the data dimensionality.

\begin{figure*}[t]
	\centering
	\includegraphics[width=0.98\textwidth]{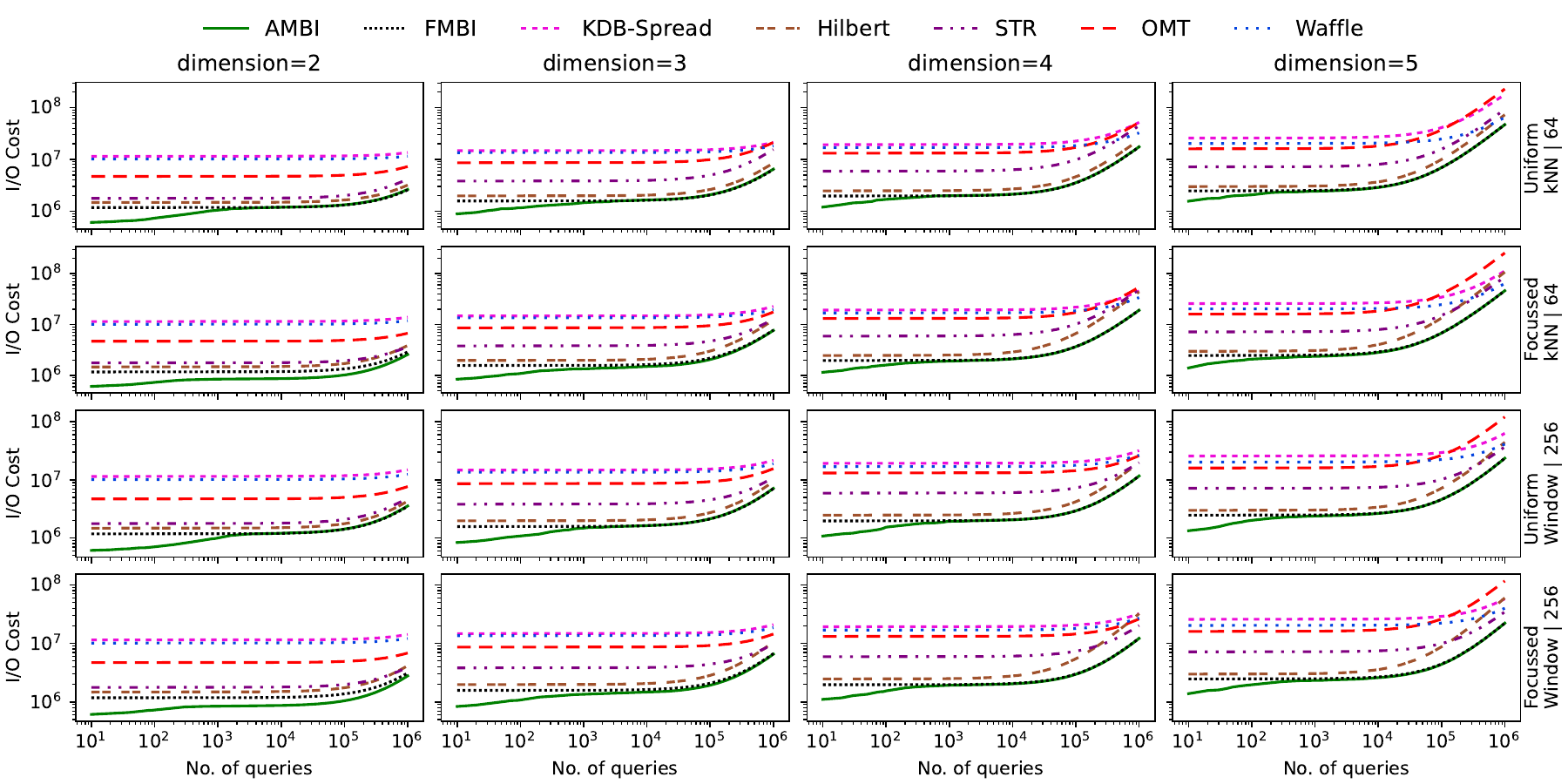}
	\caption{NYCYT | Adaptive}
	\label{fig:NYC-A}
\end{figure*}

\Cref{fig:NYC-A} focuses on the adaptive bulk loading of NYCYT. In each diagram, we measure the combined index building and cumulative query cost versus the number of queries performed, using an LRU buffer with size equal to 1\% of the dataset. Every row of diagrams corresponds to a query type and every column to a dimensionality $d$ value. For both $k$-NN and window queries, as well as for both uniform and focused\footnote{In focused queries, the query point or window lies within a 10\% volume at the center of the dataset.} distributions, the total cost increases with $d$, due to query processing; according to \Cref{fig:NYC-NA}, an average query for $d=5$ is about an order of magnitude more expensive than for $d=2$.
FMBI outperforms all non-adaptive competitors for every setting since it combines fast index building and efficient query processing. Compared to \Cref{fig:OSM-A}, AMBI converges faster to FMBI. Moreover as the dimensionality increases, the difference between uniform and focused queries diminishes due to the dimensionality curse.

\begin{figure*}[t]
	\centering
	\includegraphics[width=0.98\textwidth]{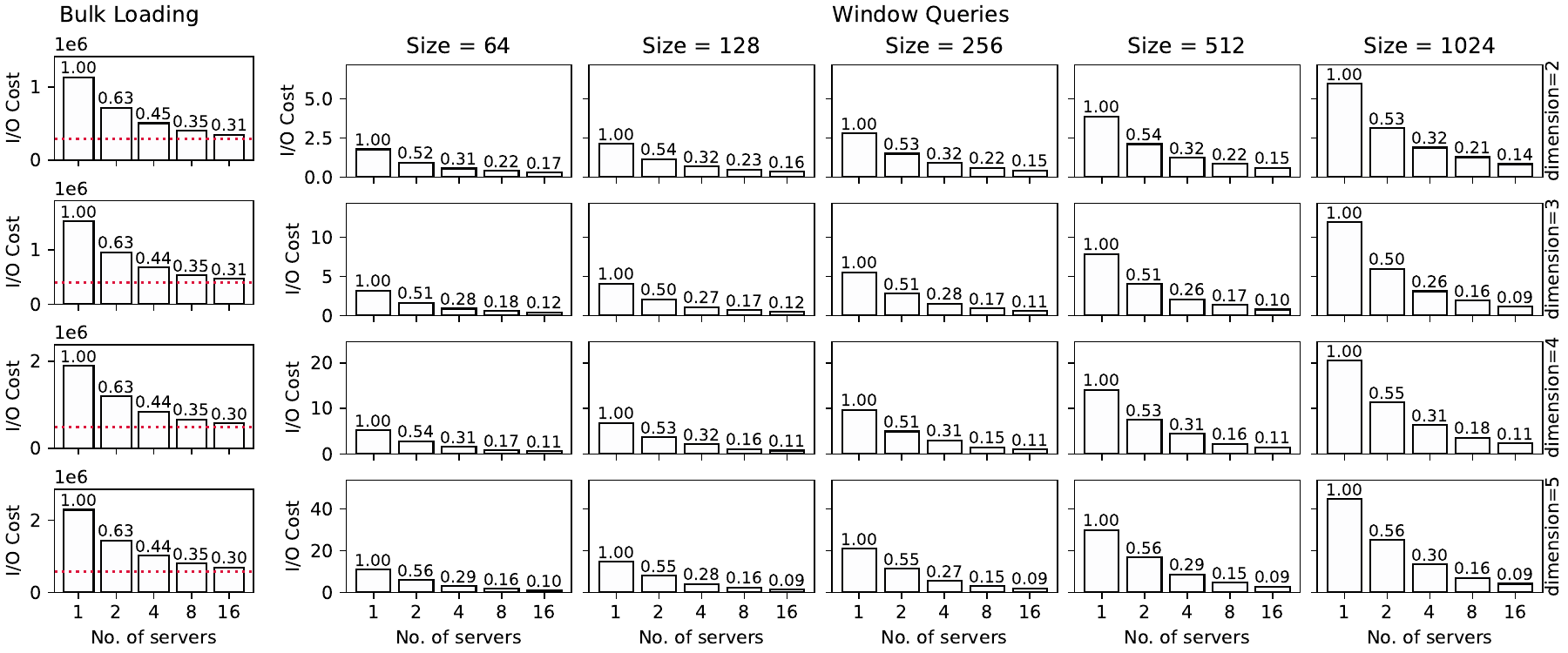}
	\caption{NYCYT | Non-Adaptive | Parallel}
	\label{fig:NYC-Par}
\end{figure*}

The final set of experiments evaluates parallel bulk loading and query processing in distributed systems with a central server and $m$ local servers, as discussed in Section 5. The leftmost diagrams in \Cref{fig:NYC-Par} illustrate the cost of building FMBI versus $m$, for NYCYT and values of $d$ between 2 and 5. The horizontal red line is the cost of scanning the entire data set at the central server, which is independent of $m$. Each server has a buffer which is equal to $\frac{5\%}{m}$ of the entire dataset. The number on top of each column indicates the relative performance with respect to a centralized architecture ($m=1$). For $m>1$, the cost is determined by the slowest local server \cite{beame_pqp_2013} ~\cite{qi_packing_2020}, i.e., the one with the densest subspace, given that all servers have identical buffers. Naturally, as the number of local servers increases, the maximum cost decreases, indicating that FMBI scales well with the number of servers because, as discussed in the context of \Cref{fig:fmbi-root-nodes} FMBI achieves well balanced partitioning. The next set of plots in \Cref{fig:NYC-Par} measures the cost of window query processing as a function of $m$. Specifically, the reported value is the number of page accesses per query, when processing 1000 uniform window queries in parallel. The parallel running time is again determined by the slowest server. As expected the cost drops as the number of servers increases. Although the actual cost grows with the dimensionality, the relative advantage of multiple local servers remains consistent.


\section{Conclusion}\label{sec:conclusion}
This paper presents novel scan-based methods for bulk loading disk-based multidimensional points.  Our first contribution is FMBI, a full index that clearly outperforms all external sort-based schemes in terms of indexing and query cost. Waffle, the only technique comparable to FMBI on query performance, is 8.2 to 8.7 times slower on index building. The adaptive version AMBI generates a partial index on demand, as a response to queries, and has substantial benefits, especially when queries are focused on a small part of the data space. Both FMBI and AMBI are naturally extended to distributed systems, where index building and query processing take advantage of multiple servers to enhance efficiency.



\clearpage

\bibliographystyle{ACM-Reference-Format}
\balance
\bibliography{ref}

\end{document}